\documentclass[lettersize,journal, 10pt]{IEEEtran}
\usepackage{amsmath,amsfonts, amssymb}
\usepackage{algorithmic}
\usepackage{algorithm}
\usepackage{array}
\usepackage{textcomp}
\usepackage{stfloats}
\usepackage{url}
\usepackage{verbatim}
\usepackage{graphicx}
\usepackage{cite}

\usepackage{hyperref}
\usepackage{listings}
\usepackage{caption}
\usepackage{bm}

\usepackage{url}

\usepackage{graphicx}%
\usepackage{multirow}%
\usepackage{mathrsfs}%
\usepackage{xcolor}%
\usepackage[table, dvipsnames]{xcolor}
\usepackage{textcomp}%
\usepackage{manyfoot}%
\usepackage{booktabs}%
\usepackage{listings}%
\usepackage{svg}
\usepackage{graphicx}
\usepackage{subcaption}
\usepackage{tabularx} 

\usepackage{amsthm}
\newtheorem{theorem}{Theorem}

\newcommand{\rev}[1]{\textcolor{black}{\ifmmode #1\else #1\fi}}

\begin{document}

\title{SparrowSNN: A Hardware/software Co-design for Energy Efficient Low Power Application}

\author{Zhenyu Bai, Zhanglu Yan*, Bin Gao, Tulika Mitra and Weng-Fai Wong 

\thanks{All authors are with the School
of Computing, National University of Singapore. Corresponding to Zhanglu Yan.  E-mail: \{zlyan, zhenyu.bai, dcstm, wongwf\}@nus.edu.sg.}% <-this % stops a space
}



\maketitle

\begin{abstract}
Deep learning has driven significant technological advancements, but its high energy consumption limits its use on battery-operated edge devices. Spiking Neural Networks (SNNs) offer promising reductions in inference-time energy consumption. 
However, existing neuromorphic architectures optimize scalable, many-core NoC execution—suited to large models but mismatched to edge devices—and their prevalent integrate-and-fire neurons re-read weights across \(T\) timesteps, inflating data-movement and dynamic-control energy.
To address this challenge, we propose SparrowSNN~\footnote{\rev{The name \emph{sparrow} comes from a Chinese proverb “Though the sparrow is small, it has all its organs”, meaning that something may be small in scale but still fully functional and well-rounded.}}, an optimized end-to-end design tailored for edge applications. SparrowSNN proposes: (1) a hardware-friendly spike activation function SSF (Sum-Spike-and-Fire); (2) a customizable $\mu$W-level-power quantized hybrid ANN-SNN model that can be designed per application; (3) a compact and low-power reconfigurable ASIC architecture, supporting the aforementioned designs. Evaluated on biomedical MIT-BIH ECG and DEAP EEG datasets, SparrowSNN achieves state-of-the-art accuracy with $20\times$ to $100\times$ lower energy consumption, significantly outperforming existing ultra-low power solutions. 
\end{abstract}

\section{Introduction}
\label{sec:introduction}
Edge devices such as portable heartbeat monitors face strict physical constraints due to their deployment scenarios. These include limited battery capacity, the need to minimize heat generation, and the requirement for long-term, uninterrupted operation. Moreover, these devices usually need to support data acquisition, on-device processing, and real-time classification, all within tight energy budgets~\cite{chen2021recent}. For such systems, energy efficiency is not merely a performance metric—it is directly tied to the practicality and usability of the device~\cite{rashid2022template,jiang2020energy}. Among various computational approaches, spiking neural networks (SNNs) stand out due to their potential for substantial energy savings. SNNs communicate through short electrical pulses, or spikes, generated only when necessary, thus leveraging event-driven and sparse computations—properties that make them particularly well-suited for energy-constrained systems~\cite{liu2024energy}.

\textbf{SNN preliminaries.}
SNNs transmit information using binary spike trains over a time window of length \(T\), consisting solely of 0s and 1s. Under rate encoding, membrane potential information is conveyed by the firing rate, i.e., the number of 1s in the spike train~\cite{stein2005neuronal,brette2015philosophy}. For instance, a value of \(\tfrac{1}{2}\) can be represented over \(T{=}4\) by an alternating spike train \([0,1,0,1]\), effectively encoding fractional values into binary sequences. Commonly, SNNs employ integrate-and-fire (IF) or leaky integrate-and-fire (LIF) models to generate such spike trains~\cite{eshraghian2023training,ganguly2020discrete,rueckauer2017conversion,wu2023dynamic,10152465}.

\begin{figure}[ht]
    \centering
    \includegraphics[width=\linewidth]{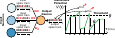}
    \caption{IF mechanism~\cite{10152465}}
    \label{fig:if_bg}
\end{figure}

In these models, spike generation is controlled by a voltage threshold \(\theta\). Using IF as an example (see Figure~\ref{fig:if_bg}): at each timestep \(t\), neuron \(i\) integrates the weighted sum of incoming $j$ spike neurons \(\bm{s}^{l-1}_j(t)\) with weights \(w^l_{ji}\) and bias \(b_i^l\) into the membrane potential \(V_i^l(t)\) per Equation~\eqref{eq:IF}. If the membrane potential crosses threshold (\(V_i^l(t)\!\ge\!\theta_i\)), the neuron fires according to Equation~\eqref{eq:spike}, emits a spike (\(s_i^l(t){=}1\)), and then resets by subtracting \(\theta_i\) from the membrane potential as in Equation~\eqref{eq:reset}; otherwise, it continues to accumulate without firing (equivalent to output 0). The LIF model fires in the same way as IF but applies an additional decay factor \(\beta\) to the previous membrane potential \(V_i^l(t{-}1)\) before the integration step in Equation~\eqref{eq:IF}.

\begin{equation}
\label{eq:IF}
V_i^l(t) = V_i^l(t-1) +\sum_j\bm{s}_j^{l-1}(t) w_{ji}^l + b_i^l
\end{equation}
\begin{equation}
\label{eq:spike}
s_i^l(t) = \begin{cases}
1, & V_i^l(t) \ge \theta_i \\
0, & \text{otherwise}
\end{cases}
\end{equation}
\begin{equation}
\label{eq:reset}
V_i^l(t) = \begin{cases}
V_i^l(t) - \theta_i, & V_i^l(t) > \theta_i \\
V_i^l(t), & \text{otherwise}
\end{cases}
\end{equation}

\begin{figure}
    \centering
    \includegraphics[width=\linewidth]{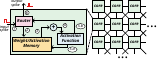}
    \caption{Conventional Digital Neuromorphic Hardware}
    \label{fig:neuromorphic}
\end{figure}

\textbf{Existing SNN inference hardware}
Most existing SNN works attribute energy gains over Artificial Neural Networks (ANNs) mainly due to two effects: (i) with binary spikes \(s\!\in\!\{0,1\}\), the multiply-accumulate between activation and weight \(s\cdot w\) is simplified to conditional additions of weigths \(w\); and (ii) event-driven execution exploits activation sparsity so idle neurons/synapses perform no work. Because general-purpose CPUs/GPUs cannot exploit this fine-grained conditionality and idleness efficiently, specialized digital neuromorphic chips—most notably IBM TrueNorth~\cite{akopyan2015truenorth} and Intel Loihi~\cite{lines2018loihi}—have been developed for SNN inference. As shown in Figure~\ref{fig:neuromorphic}, these systems use a 2D mesh of small cores. Each core keeps the weights and the neuron state in their local SRAM. Spikes arrive over the Network-on-Chip (NoC) \textcircled{1} and trigger an event-driven pipeline: fetch the relevant synaptic weights \textcircled{2}; read the destination neurons’ membrane potentials (activations) \textcircled{3}; accumulate inputs and evaluate the firing threshold to generate output spikes \textcircled{4}; write back updated membrane potentials \textcircled{5.a}; and inject any generated spikes into the router for delivery to downstream cores \textcircled{5.b},\textcircled{6}.

\textbf{Limitations.}
Most neuromorphic systems target scalability: many small cores connected by a NoC, well suited to large workloads. To keep utilization high, multiple neurons (and their synapses) are mapped to one core. With IF/LIF neurons, execution is event- (spike-) driven: each arriving spike triggers a fetch of the corresponding synaptic weight(s), a read–modify–write of the destination membrane potential, and a threshold/reset check—repeated \textbf{\textit{per spike}} over a window of \(T\) timesteps; LIF further incurs periodic leakage updates even without spikes. The resulting memory traffic and fine-grained control flow inflate data-movement and dynamic-control energy~\cite{akopyan2015truenorth,tang2023hardware,bhattacharjee2024snns,chu2022neuromorphic}. In addition, the scale of these many-core fabrics typically keeps power in at least the watt range, which is mismatched to the \(\mu\)W–mW budgets of battery-powered edge devices.

\textbf{Our contribution 1: New SNN neuron model, SSF. }
To eliminate the per-timestep reloading of weights and neuron state, we introduce \emph{sum-spikes-and-fire} (SSF)—a hardware-friendly spike-generation primitive that replaces the \(T\)-step IF/LIF loop with a single accumulation over the window followed by one threshold decision. This removes inter-timestep dependencies and avoids redundant memory traffic. SSF is not only more power efficient but has a supperior representability compared to classical IF-based neuron model.

\textbf{Our contribution 2: Low-power ASIC for hybrid ANN/SNN inference.}
We then designed a reconfigurable ASIC design to support our proposed SSF firing mechanism. The architecture tightly couples compute with on-chip SRAMs and uses simplified, deterministic control to curb data movement and dynamic-control energy. It natively supports SSF and remains backward-compatible with IF neurons, quantized ANNs, and their hybrids, enabling seamless deployment across model families under low-power edge constraints. Empirically, in our experiments, SSF provides the lowest energy, but is alone insufficient for robust feature extraction on raw signals. We found that placing one or more quantized ANN layers at the front end for feature extraction, followed by SSF-based SNN layers, yielding an improved accuracy–energy balance while preserving low-power operation.

\textbf{Our contribution 3: End-to-end model/hardware co-design.}
Because quantized ANNs and SSF-based SNNs trade off accuracy and energy differently, we introduce an automated framework that selects a model structure and generates hardware configuration to meet a target model accuracy for a given application—up to per-patient personalization, while minimizing the energy consumption. The design workflow is:
\begin{enumerate}
    \item encode the target dataset or per-patient data (e.g., MIT-BIH for electrocardiogram (ECG), DEAP for electroencephalogram(EEG));
    \item run \emph{Network Architecture Search} (NAS) to find the appropriate network structure with respect to the system requirements on accuracy, energy, and performance;
    \item choose the neuron type per layer—quantized ANN, SNN with IF, or SSF—and assemble a hybrid model; and
    \item deploy the model by reconfiguring our ASIC for inference.
\end{enumerate}

On ECG (MIT-BIH), our system attains \textbf{98.61\,\%} accuracy at \textbf{11.76\,nJ} per inference. On EEG (DEAP, arousal), it achieves \textbf{85.31\,\%} at \textbf{17.87\,nJ} per inference. 
% Our key contributions are:

% \begin{itemize}
% \item \textbf{Hardware-optimized spike generation:} \emph{Sum-Spikes-and-Fire (SSF)} reduces memory accesses and removes inter-timestep dependencies compared to IF/LIF.
% \item \textbf{$\mu$W-level hybrid ANN–SNN model:} Quantized ANN front-end for robust feature extraction followed by SSF-based SNN layers, delivering state-of-the-art accuracy on MIT-BIH and DEAP at ultra-low power.
% \item \textbf{Reconfigurable single-core ASIC for edge deployment:} A compact, programmable chip that natively supports SSF, IF neurons, quantized ANNs, and hybrids, cutting inference energy while preserving accuracy.
% \item \textbf{End-to-end deployment workflow:} NAS-driven model selection, neuron-type assignment, and direct hardware mapping tailored to edge applications.
% \end{itemize}

\section{Improving IF with SSF Spiking}
\begin{figure}[h]
    \centering
    \includegraphics[width=.5\linewidth]{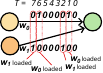}
    \caption{Typicall mapping of IF neurons, causing multiple weight loading}
    \label{fig:redundant-weight-loading}
\end{figure}

SNNs naturally exhibit sparsity—most entries in a spike train are zeros—so neuromorphic chips process spikes in an \emph{event-driven} manner. In a classical IF/LIF pipeline, each arriving spike carries indices for source and destination neurons and \emph{triggers} memory activity on a core that holds a group of neurons: fetch synaptic weight(s), read/modify/write the destination membrane potential, evaluate threshold/reset, and, if needed, emit a new spike downstream. While this exploits activation sparsity, it also causes \emph{redundant} SRAM traffic: \textbf{every spike} reloads the same weight(s) and membrane state. As illustrated in Figure~\ref{fig:redundant-weight-loading}, repeated per-spike accesses dominate energy.

\subsection{Why per-spike IF is expensive.}

Table~\ref{tab:energy-comp} reports the energies (synthesized with standard cells and memory synthesis tools at 22\,nm, TT process, 0.85V, 25C). A single 8b\(\rightarrow\)16b integer accumulation (ACC) costs \(\approx0.05\) pJ, versus \(\approx0.13\) pJ for an 8b\(\times\)8b\(\rightarrow\)16b Multiply-and-Accumulation (MAC). 
\begin{table}[h]
    \centering
    \begin{tabular}{c|c}
         8b-16b INT Acc &  0.05 pJ  \\
         8b-8b-16b INT MUL & 0.1 pJ \\
         8b-8b-16b INT MAC & 0.13 pJ \\
         %16b-16b-32b INT MUL & ? fJ  \\ %optional
         \hline
         4KB SRAM 128b read & $\approx$3 pJ (0.18 pJ/Byte)  \\
         4KB SRAM 128b write & $\approx$5 pJ (0.31 pJ/Byte) \\
         64KB SRAM 128b read& $\approx$4 pJ (0.25 pJ/Byte)\\ 
         64KB SRAM 128b write& $\approx$8 pJ (0.5 pJ/Byte)\\ 
    \end{tabular}
    \caption{Energy Consumption of different operations at a commercialized 22nm technology node. Compute operation results are obtained from synthesizing default RTL library with standard cells. Memory results are obtained from memory synthesis tool with low-power configurations. }
    \label{tab:energy-comp}
\end{table}

Therefore, purely from the energy consumption of \emph{compute},
an IF update (just an ACC when \(s\in\{0,1\}\)) beats a MAC-based ANN only when the average number of generated spikes per input over a window, \(s\), is small:
\begin{equation}
\label{eq:compute_only}
E_{\text{compute}}^{\text{IF}} = s\,E_{\text{ACC}},\quad
E_{\text{compute}}^{\text{ANN}} = E_{\text{MAC}}
\end{equation}
\begin{equation}
\label{eq:compute_cross}
E_{\text{compute}}^{\text{IF}}\le E_{\text{compute}}^{\text{ANN}}
\;\Leftrightarrow\;
s \le \frac{E_{\text{MAC}}}{E_{\text{ACC}}}
\approx \frac{0.13}{0.05} \approx 2.6.
\end{equation}

However, the memory operations are much more expensive. Reading \(\!1\) byte from a 64\,KB SRAM costs \(\approx0.25\) pJ; reading/writing \(\!1\) byte in a 4\,KB SRAM costs \(\approx0.18/0.31\) pJ (Table~\ref{tab:energy-comp}). If each spike performs \(\textbf{(i)}\) one-byte weight (assuming 8b weight quantization) read from 64\,KB, \(\textbf{(ii)}\) one-byte membrane (activation) read from 4\,KB, and \(\textbf{(iii)}\) one-byte membrane write, then with \(s\) spikes per window the memory energies are:
\begin{equation}
\label{eq:mem_only}
\begin{split}
E_{\text{memory}}^{\text{IF}} = s \cdot E_{\text{read}}^{w} + \max(s-1,0)\left(E_{\text{read}}^{V}+E_{\text{write}}^{V}\right),\\
E_{\text{memory}}^{\text{ANN}} = E_{\text{read}}^{w}
\end{split}
\end{equation}
where the \(\max(s-1,0)\) term avoids charging a read/write when \(s{=}0\) and counts one fewer read-modify-write operation on the first event.\footnote{This is a conservative simplification; alternative pipelines change the constant terms but not the conclusion.}

Adding compute and memory together gives the crossover:
\begin{gather}
\label{eq:total_if}
E^{\text{IF}} = s\,E_{\text{ACC}} + s\,E_{\text{read}}^{w} + \max(s-1,0)\left(E_{\text{read}}^{V}+E_{\text{write}}^{V}\right),\\
\label{eq:total_ann}
E^{\text{ANN}} = E_{\text{MAC}} + E_{\text{read}}^{w} \rev{+ E^{activation}_{read} + E^{activation}_{write}},
\end{gather}
\begin{gather}
\label{eq:mem_cross}
E^{\text{IF}}\le E^{\text{ANN}}
\;\Rightarrow\;
s \leq \\ \frac{E_{\text{MAC}} + E_{\text{read}}^{w} + E_{\text{read}}^{V}+E_{\text{write}}^{V}\rev{+ E^{activation}_{read} + E^{activation}_{write}}} 
{E_{\text{ACC}} + E_{\text{read}}^{w} + E_{\text{read}}^{V}+E_{\text{write}}^{V}}
%\approx \frac{0.13{+}0.25{+}0.18{+}0.31}{0.05{+}0.25{+}0.18{+}0.31}
\end{gather}
Thus IF only wins when the \emph{average} spike count per input per window is \rev{\(\lesssim1.7\)}. In our later evaluation on MIT-BIH dataset for ECG classification, an IF-SNN meets this condition only at \rev{\(T{=}7\), which reduces its accuracy by approximately 12\% (Sec.~\ref{sec:result})}.

\subsection{Sum-Spikes-and-Fire (SSF)}
The energy analysis above shows that per-spike read–modify–write of membrane state and weights is the dominant cost. Consequently, the path to lower energy is to remove per-spike updates while preserving the information that matters. We make a key observation under the rate encoding methods (including IF), that the update of Equation~\eqref{eq:IF}, the total integrated input over a window depends on \emph{how many} spikes occurred, not \emph{when} they occurred. Therefore, the timing within the window is redundant, and we can summarize the presynaptic activity by a spike \emph{count}:

\begin{equation}
\label{eq:ssf_count}
c^{\,l-1}_j \;=\; \sum_{t=1}^{T} s^{\,l-1}_j(t) \;\in\; \{0,\ldots,T\}.
\end{equation}

Building on this, SSF replaces \(T\) per-spike updates with a \emph{single} per-window computation. \emph{First}, we integrate once using the counts (bias scales with \(T\)):
\begin{equation}
\label{eq:ssf_integrate}
u^l_i \;=\; \sum_{j} c^{\,l-1}_j\, w^l_{ji} \;+\; T\,b^l_i,
\end{equation}
which is algebraically equivalent to summing Equation~\eqref{eq:IF} over \(t{=}1\ldots T\). \emph{Then}, we convert this integrated value to an \emph{output count} consistent with IF’s subtractive reset:
\begin{equation}
\label{eq:ssf_fire}
c^{\,l}_i \;=\; \left\lfloor \frac{u^l_i}{\theta_i} \right\rfloor \quad \text{(optionally clipped to }[0,T]\text{)}.
\end{equation}

\begin{figure}[ht]
    \centering
    \includegraphics[width=0.95\linewidth]{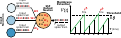}
    \caption{Example of neuron firing with SSF mechanism}
    \label{fig:ssf}
\end{figure}

In fact, as shown in the Algorithm~\ref{alg:ssf} and Figure~\ref{fig:ssf}, SSF actually decouples \emph{integrate} and \emph{fire}: (i) count presynaptic spikes once, (ii) perform one dot product per neuron via Equation~\eqref{eq:ssf_integrate}, and (iii) map to an output count via Equation~\eqref{eq:ssf_fire}. \emph{As a result}, SSF removes inter-timestep dependencies and cuts weight/membrane SRAM traffic by up to \(T\times\). The details of the hardware implementaion leveraging the advantage of the SSF mechanism and further optimizations are shown in the next section (Section~\ref{sec:hardware_design}).

% We replace a \(T\)-bit spike train by a count:
% \begin{equation}
% \label{eq:ssf_count}
% c^{\,l-1}_j \;=\; \sum_{t=1}^{T} s^{\,l-1}_j(t) \;\in\; \{0,\ldots,T\},
% \end{equation}
% encoded in \(\lceil\log_2(T{+}1)\rceil\) bits. Using these counts, neuron \(i\) performs a \emph{single} accumulation for the window:
% \begin{equation}
% \label{eq:ssf_integrate}
% u^l_i \;=\; \sum_{j} c^{\,l-1}_j\, w^l_{ji} \;+\; T\,b^l_i,
% \end{equation}
% which is algebraically equivalent to summing Equation~\eqref{eq:IF} over \(t{=}1\ldots T\) with rate encoding. The output is then a \emph{count} of spikes, consistent with IF’s subtractive reset:
% \begin{equation}
% \label{eq:ssf_fire}
% c^{\,l}_i \;=\; \left\lfloor \frac{u^l_i}{\theta_i} \right\rfloor \;\;\; \text{(optionally clipped to }[0,T]\text{)}.
% \end{equation}
% Equivalently, a binary “fired-at-least-once” decision uses \(1[u^l_i \ge T\theta_i]\).
% Operationally, SSF (Figure~\ref{fig:ssf}) decouples \emph{integrate} and \emph{fire}: 
% \(\,\)(i) count spikes once per input, 
% \(\,\)(ii) do one dot-product per neuron (Equation~\eqref{eq:ssf_integrate}), 
% \(\,\)(iii) convert to an output count via Equation~\eqref{eq:ssf_fire}. 
% This converts IF’s per-spike state updates into a single per-window operation, cutting weight/membrane SRAM traffic by up to \(T\!\times\) and removing inter-timestep dependencies. Hardware details appear in Section~\ref{sec:hardware_design}.

\begin{algorithm}[ht]
\caption{Sum-Spikes-Fire Model}
\label{alg:ssf}
\begin{algorithmic}[1]
\REQUIRE Time window size $T$, weights $\bm{w}_{ji}^{l}$, bias $b_i^{l}$, input spike trains $\{s^{l-1}_j(t)\}_{t=1}^T$ for layer $l$ and neuron $j$ with \(|s_j^l|\) number of spikes, threshold $\theta$
\ENSURE Output spike number $|s_i^{l}|$ of neuron $i$ for layer $l$
\STATE \textbf{Step 1: Sum Spikes}
\STATE Initialize cumulative spike input: $S_i^{l} \leftarrow 0$
%\FOR{$k=1$ \TO $|s_j^{(l)}|$}
    \STATE $S_i^{l} \leftarrow \sum_j\bm{w_{ji}}^{l}(t) |s_j^{(l)}| + b_i^{l}$
%\ENDFOR

\STATE \textbf{Step 2: Fire}
\STATE Initialize membrane potential: $V_i^{l} \leftarrow 0$
\FOR{$t=1,2,\dots,T$}
    \STATE Update membrane potential: $V_i^{l} \leftarrow V_i^{l} + S_i^{l}$
    \IF{$V_i^{l} \geq T\theta_i$}
        \STATE Emit spike: $s_i^{l}(t) \leftarrow 1$; \(|s_i^{l}|+=1\)
        \STATE Reset membrane potential: $V_i^{l} \leftarrow V_i^{l} - T\theta$
    \ENDIF
\ENDFOR
\RETURN $|s_i^{l}|$
\end{algorithmic}
\end{algorithm}

\subsection{Representability of SSF}
\begin{figure}[h]
\begin{subfigure}{.49\linewidth}
    \includegraphics[width=\linewidth]{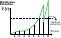}
    \caption{IF overflow}
    \label{fig:IF-overflow}
\end{subfigure}
\begin{subfigure}{.49\linewidth}
    \includegraphics[width=\linewidth]{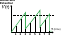}
    \caption{SSF prevents overflow}
    \label{fig:SSF-overflow}
\end{subfigure}
\end{figure}

Classical IF encoding can undercount the spikes when presynaptic spikes (especially large-weight ones) \emph{cluster late} in the time window. In IF with subtractive reset, each emitted spike removes only one threshold \(\theta\) from the membrane; if several threshold crossings accrue near the window end, the neuron can emit at most one spike per timestep and any remaining “backlog” stays as residual potential \(V(T)\ge\theta\), which cannot be discharged once the window closes. This \emph{overflow} effect, as shown in Figure~\ref{fig:IF-overflow}, loses information. In contrast, SSF first integrates all contributions in the window and then converts the total to an \emph{output count} (Equation~\eqref{eq:ssf_integrate}–\eqref{eq:ssf_fire}), so late clustering does not reduce the reported spike count (Figure~\ref{fig:SSF-overflow}). In typical cases where the final residual is \(<\theta\), SSF and IF produce the same count; in adversarial late-clustered cases, SSF is strictly more representative of the integrated input.

\begin{theorem}[SSF representability vs.\ IF]
\label{thm:ssf-repr}
Consider a single neuron with subtractive-reset threshold \(\theta>0\) over a window of \(T\) timesteps. Let the per-timestep input be
\[
x(t)=\sum_j w_{ji}\,s_j(t)+b,\quad t=1,\ldots,T,
\]
with \(s_j(t)\in\{0,1\}\) and \(w_{ji},b\ge 0\). Define the integrated input \(U=\sum_{t=1}^{T}x(t)\).
\begin{enumerate}
\item \textbf{SSF count.} The SSF output count is timing-invariant and equals
\[
N_{\mathrm{SSF}}=\min\!\Big\{T,\,\big\lfloor U/\theta \big\rfloor\Big\}.
\]
\item \textbf{IF vs.\ SSF.} Let \(N_{\mathrm{IF}}\) be the number of spikes emitted by discrete-time IF with at most one spike per timestep. Then
\[
N_{\mathrm{IF}} \;\le\; N_{\mathrm{SSF}},
\]
with equality if and only if the cumulative sum \(S(t)=\sum_{\tau=1}^{t}x(\tau)\) crosses the levels \(\theta,2\theta,\ldots, N_{\mathrm{SSF}}\theta\) at \(N_{\mathrm{SSF}}\) (not necessarily consecutive) timesteps within the window.%\footnote{A sufficient (but not necessary) condition for equality is the per-timestep bound \(x(t)\le\theta\) for all \(t\), which ensures the rest of membrane potential after firing is never above the threshold. multi-threshold backlogs within one step.}
\end{enumerate}
\end{theorem}

\begin{proof}
\emph{(1) SSF.)} By Eq.~\eqref{eq:ssf_integrate}, SSF computes \(u_i^l=\sum_j c^{\,l-1}_j w_{ji}+T b = \sum_{t=1}^{T}x(t) = U\) and then outputs \(c_i^l=\lfloor u_i^l/\theta\rfloor\), optionally clipped to \([0,T]\) (Eq.~\eqref{eq:ssf_fire}). Hence \(N_{\mathrm{SSF}}=\min\{T,\lfloor U/\theta\rfloor\}\), which depends only on \(U\), not on the timing of \(x(t)\).

\emph{(2) IF vs.\ SSF.)} For IF, let \(S(t)=\sum_{\tau=1}^{t}x(\tau)\) be the cumulative (no-reset) potential and \(N_{\mathrm{IF}}(t)\) the number of emitted spikes up to \(t\). The actual membrane satisfies
\[
V(t)=S(t)-\theta\,N_{\mathrm{IF}}(t),\qquad V(t)\in[0,\theta) \;\; \text{by subtractive reset}.
\]
Because IF emits \(\le 1\) spike per timestep, \(N_{\mathrm{IF}}(T)\le \lfloor U/\theta\rfloor\). If \(\lfloor U/\theta\rfloor>T\) then the per-timestep cap further implies \(N_{\mathrm{IF}}(T)\le T\). Combining these gives
\[
N_{\mathrm{IF}}(T)\le \min\{T,\lfloor U/\theta\rfloor\}=N_{\mathrm{SSF}}.
\]
For equality, IF must fire exactly once for each multiple of \(\theta\) reached by the cumulative sum, before the window ends. That is, for every \(k\in\{1,\ldots,N_{\mathrm{SSF}}\}\) there must exist a timestep \(t_k\le T\) such that \(S(t_k)\ge k\theta\) and the spikes are realized at (some of) these steps (one per step). If any two or more multiples are crossed within the same timestep (i.e., \(S(t)-S(t-1)>\theta\)) or too late to realize the full backlog before \(T\), then IF cannot emit all \(N_{\mathrm{SSF}}\) spikes and strict inequality holds. The stated level-crossing condition is therefore necessary and sufficient.
\end{proof}

\noindent Part~(1) formalizes that SSF returns the \emph{ideal} count implied by total integrated input (clipped by the obvious limit of \(T\) spikes). Part~(2) shows discrete-time IF can only match SSF when threshold crossings are sufficiently separated in time; late clustering or large single-step increments create a backlog that cannot be discharged within the window, leading to undercounting in IF but not in SSF. Practically, this is exactly the overflow scenario in Figure~\ref{fig:IF-overflow} that SSF corrects as shown in Figure~\ref{fig:SSF-overflow}.

\section{Tuning Model Architecture for Optimal Performance–Energy Tradeoff}
\begin{figure}[h]
    \centering
    \includegraphics[width=\linewidth]{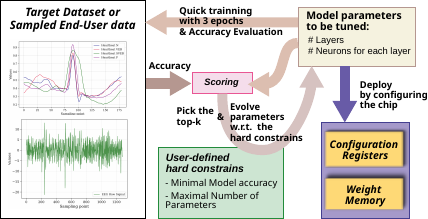}
    \caption{SparrowSNN Workflow}
    \label{fig:workflow}
\end{figure}

Different end-user applications impose different targets on accuracy, energy, and latency. For example, medical monitoring often has strict latency requirements (set by the sampling and inference rate), stringent accuracy criteria for certification, and tight power budgets for battery-powered or implantable devices. To expose explicit accuracy–energy–latency trade-offs, we introduce a framework that tunes the network architecture to a user-specified operating point and returns a Pareto-optimal configuration \rev{for our custom ASIC design with respect to the corresponding performance/power model}. Figure~\ref{fig:workflow} summarizes the workflow. Because SNNs are hard to train, we use a quantized ANN as a proxy during search and then convert the selected architecture to its SNN/SSF counterpart for deployment. The tunable parameters are the number of layers and the number of neurons per layer. These map directly to our reconfigurable ASIC (Section~\ref{sec:hardware_design}) via configuration registers (e.g., layer count and types) and \emph{Weight Memory} (loaded parameters). To respect hardware limits, the search is restricted to architectures with 3–6 layers and 16–128 neurons per layer (powers of two). 

\begin{figure}[h]
    \centering
    \includegraphics[width=\linewidth]{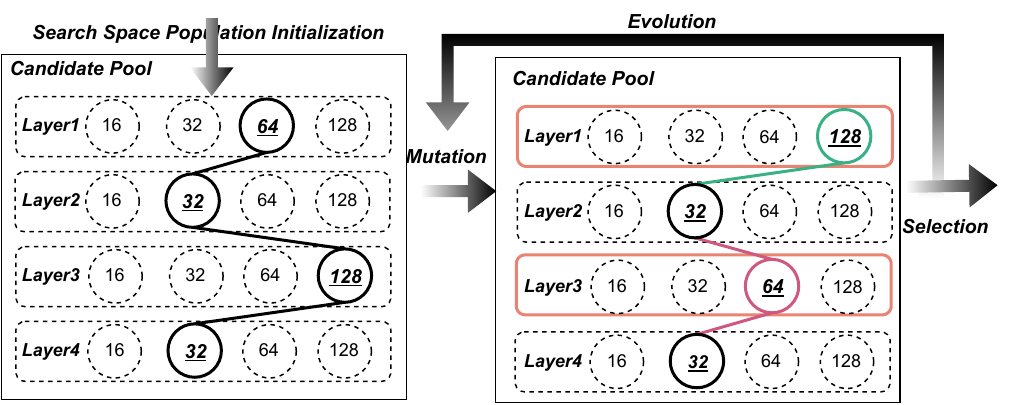}
    \caption{Network Architecture Search }
    \label{fig:nas}
\end{figure}

\noindent\textbf{Search procedure.}
Because the design space is large, we employ an \emph{evolutionary} search (Figure~\ref{fig:nas}). \rev{While the search methodology is similar to existing NAS works~\cite{snn-nas-1,snn-nas-2,snn-nas-3}, our search space is strictly bounded to ensure that all candidate architectures can be mapped to our custom ASIC, and are among the pareto-optimal points with respect to the performance/power model.} We initialize 500 random candidate architectures and, for each, perform a brief training run of 3 epochs on the target dataset or sampled end-user data to estimate accuracy. The accuracy on the target data together with the model architecture are used for scoring each candidate based on a user-defined objective function. The top candidates that satisfy the user-defined hard constrains (typically minimal model accuracy and maximal number of parameters) are selected as parents; their layer widths are mutated with probability 0.2 to form the next generation~\cite{winter2025ecological}. 
\rev{We iterate selection–mutation–brief retraining for 50 rounds, and then train the best-scoring model for 150 epochs before converting and finally deploying it. Our search space consists of networks with $L$ layers ($L \in \{3, 4, 5, 6\}$), where each layer independently takes one of 4 dimensions ($D \in \{16, 32, 64, 128\}$) which is a common setting for small network structure. The total number of possible architectures is:$$|\mathcal{S}| = \sum_{L=3}^{6} |D|^L = 4^3 + 4^4 + 4^5 + 4^6 = 5440$$.
The overall time complexity is 
$
\mathcal{O}((N_{init} + R \cdot K_{best} \cdot M) \cdot (E_{proxy} \cdot K_{folds}))
$, where the parameters are defined as follows: $N_{init} = 500$ is the number of initial random models, $R = 50$ is the number of mutation rounds, $K_{best} = 3$ represents the top models selected as parents in each round, $M = 10$ is the number of mutant offspring generated per parent, $E_{proxy} = 3$ denotes the proxy epochs used for quick evaluation, and $K_{folds} = 3$ is the number of cross-validation folds.}

With the NAS-selected macro-architecture in hand, we instantiate and compare three neuron types: quantized ANN, IF-based SNN, and our SSF-based SNN. Across evaluations, SSF consistently outperforms IF at matched spike-window length \(T\); for example, on DEAP (EEG) valence classification with \(T{=}6\), SSF improves accuracy by up to \(\mathbf{20.45\%}\) over IF.

However, purely SSF-based models can struggle to capture fine-grained features when \(T\) is small and inputs are low-bit quantized. To address this, we adopt a \emph{Hybrid ANN–SSF} design: quantized ANN layers at the front end for robust feature extraction, followed by SSF layers for low-energy inference. Conceptually, ANN and SSF represent complementary points on the accuracy–energy spectrum—ANN favoring accuracy in early feature formation, SSF favoring energy efficiency in later processing. This hybridization preserves accuracy at low \(T\) while keeping energy low. On DEAP, the Hybrid model achieves state-of-the-art results—\(\mathbf{85.04\%}\) (valence) and \(\mathbf{85.31\%}\) (arousal) at \(T{=}31\)—exceeding the pure SSF model by \(\sim\)2–3\% with only a modest increase in energy. Power can be further reduced by low-bit quantization of both ANN and SNN weights/activations, as described in Algorithm~\ref{alg:quant_infer}.

%Algorithm~\ref{alg_q_q} is applied on the trained ANN $M$ to perform weight and activation quantization which serving as the encoding layer for information extraction . Initially, we determine the maximum and minimum values of both the weights and biases, \(f_{\text{max}}\) and \(f_{\text{min}}\), to compute the scaling factor \(s_w\) for weight and bias quantization. We then run \(M\) on the entire training dataset to gather the scaling factors \(s_i\) for input activations and \(s_o\) for output activations, based on the layer's maximum and minimum values across all training data. During quantized ANN inference, input data is scaled to its integer representation, multiplied by the integer weights, and added to the integer biases. Rescaling factors \(r_1\) and \(r_2\) map the integer output \(x_{o,q}\) back to its original range. %This process is detailed in Algorithm~\ref{alg_q_q}.

\begin{algorithm}[ht]
\caption{Quantized Inference }
\label{alg:quant_infer}
\begin{algorithmic}[1]
\REQUIRE  Weights $\bm{w}^{l}$ and bias $b^{l}$ of layer $l$, quantization bit-width $q$, training dataset, shifting bits $n_{\text{shift}}$ and bias shifting bits $m_{\text{shift}}$.
\ENSURE Quantized inference output $\bm{x}^{l+1}_{q}$ for layer $l$. Out spike train $\{s^{l+1}(t)\}_{t=1}^T$
\STATE \textbf{Collect statistics over the training dataset}
\STATE Compute scale for weights and biases:
\STATE $r_w \leftarrow \frac{\max\{\bm{w}^{l}, b^{l}\} - \min\{\bm{w}^{l}, b^{l}\}}{2^{q}-1}$
\STATE Quantize weights and biases:
\STATE $\bm{w}^{l}_q \leftarrow \text{clamp}\left(\left\lfloor\frac{\bm{w}^{l}}{r_w}\right\rceil, -2^{q-1}, 2^{q-1}-1\right)$
\STATE $b^{l}_q \leftarrow \text{clamp}\left(\left\lfloor\frac{b^{l}}{r_w}\right\rceil, -2^{q-1}, 2^{q-1}-1\right)$

\STATE For ANN layer, collect input/output scales :
\STATE $r_i \leftarrow \frac{x^l_{\text{max}} - x^l_{\text{min}}}{2^{q}-1}$, $r_o \leftarrow \frac{x^{l+1}_{\text{max}} - x^{l+1}_{\text{min}}}{2^{q}-1}$
\STATE $r_w \leftarrow \left\lfloor (r_i*r_w/r_o \times 2^{n_{\text{shift}}})\right\rceil $ , $r_b \leftarrow \left\lfloor(r_w/r_o \times 2^{m_{\text{shift}}})\right\rceil$

\STATE For SNN layer, quantize threshold :
\STATE $\theta_q \leftarrow \left\lfloor\frac{\theta}{r_w}\right\rceil$

\STATE \textbf{Perform Quantized SNN  Inference}
\STATE Send $\bm{w}^{l}_q,b^{l}_q, \theta_q$ and $\bm{s^l}$ to SSF in Algo1.
\RETURN $\{s^{l+1}(t)\}_{t=1}^T$

\STATE \textbf{Perform Quantized ANN  Inference}
\STATE $\bm{x}^{l+1}_{q} \leftarrow \left\lfloor \bm{w}_q^{l}\bm{x}_{i,q}  \right\rfloor *r_w \gg n_{shift}+ r_b*\left\lfloor b_q^{l}  \right\rfloor \gg m_{shift}$
\STATE $\bm{x}^{l+1}_{q} \leftarrow \text{clamp}\left(\bm{x}^{l+1}_{q} ,0,2^{q}-1,\right)$
\RETURN $\bm{x}^{l+1}_{q}$

\end{algorithmic}
\end{algorithm}

\section{SparrowSNN ASIC design}
\label{sec:hardware_design}
Targeting wearable and implantable devices, SparrowSNN must operate under power budgets orders of magnitude below watt-level neuromorphic chips. Achieving \(\mu\)W-class operation requires carefully co-designed control, memory, and datapaths to minimize switching activity and memory traffic. This section details the design challenges and the architectural choices that enable a reconfigurable, ultra–low-power implementation.

\begin{figure}[h]
    \centering
    \includegraphics[width=\linewidth]{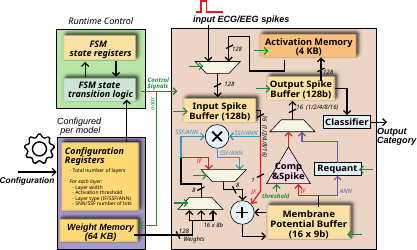}
    \caption{Hardware Design}
    \label{fig:hardware-design}
\end{figure}

\paragraph{Architecture overview (Fig.~\ref{fig:hardware-design}).}
SparrowSNN consists of three blocks: (1) \emph{configuration registers and weight memory}, (2) a computation core with a flexible datapath and local buffers, and (3) an \emph{FSM-based runtime controller}. Edge workloads of interest are compact enough to fit a single core; larger models cannot satisfy real-time latency within a \(\mu\)W budget. At \(100\)~MHz, SparrowSNN supports \(\sim 60\)K parameters (matching the weight memory) and sustains \(\sim\)180 inferences/s (180~Hz).

Configuration registers store model hyperparameters (number of layers; per-layer widths; neuron types), while the weight memory holds all weights and biases. These are programmed once per model and remain static in the field.

The FSM controller sequences inference by enabling different units based on the configured layer/width/type. It tracks layer/neuron indices, generates layer-transition signals, and computes addresses for weight and activation fetches. The core supports up to six layers, each with up to 128 input and 128 output neurons. Each layer can operate as IF, SSF, or (quantized) ANN.

\subsection{Hardware Synthesis}
We synthesized SparrowSNN in a commercial 22\,nm process using Synopsys Design Compiler; SRAMs were generated with dedicated memory compilers in low-power configurations. The evaluated corner is at typical (TT) process, 0.85V, 25C temperature.  The total chip area is \(114{,}689.96\,\mu\text{m}^2\), with \(\sim\)90\% weight memory, 7.2\% activation memory, and 2.7\% compute+control. At 0.85\,V and 100\,MHz, the worst-case (toggle-rich) power is \(\sim\)60\,\(\mu\)W; actual power depends on the configured network and activity. The 100\,MHz/0.85\,V operating point is chosen for our EEG/ECG showcases; smaller models admit  Dynamic Voltage and Frequency Scaling (DVFS) to further reduce dynamic and leakage power.

\subsection{Datapath Design}
ANN, IF, and SSF share most of the datapath (black in Figure~\ref{fig:hardware-design}); only a few mode-specific blocks differ (blue: SSF/ANN; red: IF; purple: ANN-only). All modes use the same weight bit-width to simplify model loading procedure and reduce control complexity.

Activation values are either provided externally as inputs to the first layer or stored internally between layers in dedicated activation memory (SRAM). This dedicated activation memory is optimized for reduced energy cost per memory operation due to its smaller size relative to the weight memories. 

%\textit{Two-level memory with width adaptation.} Memory operations dominate energy; thus, SparrowSNN uses wide SRAM ports with local buffering to serve fine-grained consumers efficiently. Figure~\ref{fig:mem-portwidth} (normalized) shows energy/bit vs.\ port width: wider ports reduce energy/bit but require buffers to down-convert to the compute granularity. Concretely, both weight and activation SRAMs use 128-bit ports. Weights are consumed as 8-bit values; activations/spike representations are consumed at 1\,bit (IF) or 2/4/8/16\,bits (SSF or ANN), selected per layer. Local input/output buffers bridge the 128-bit SRAM bursts to these per-op widths, minimizing total memory energy while keeping utilization high.

\begin{figure}
    \centering
    \includegraphics[width=.9\linewidth]{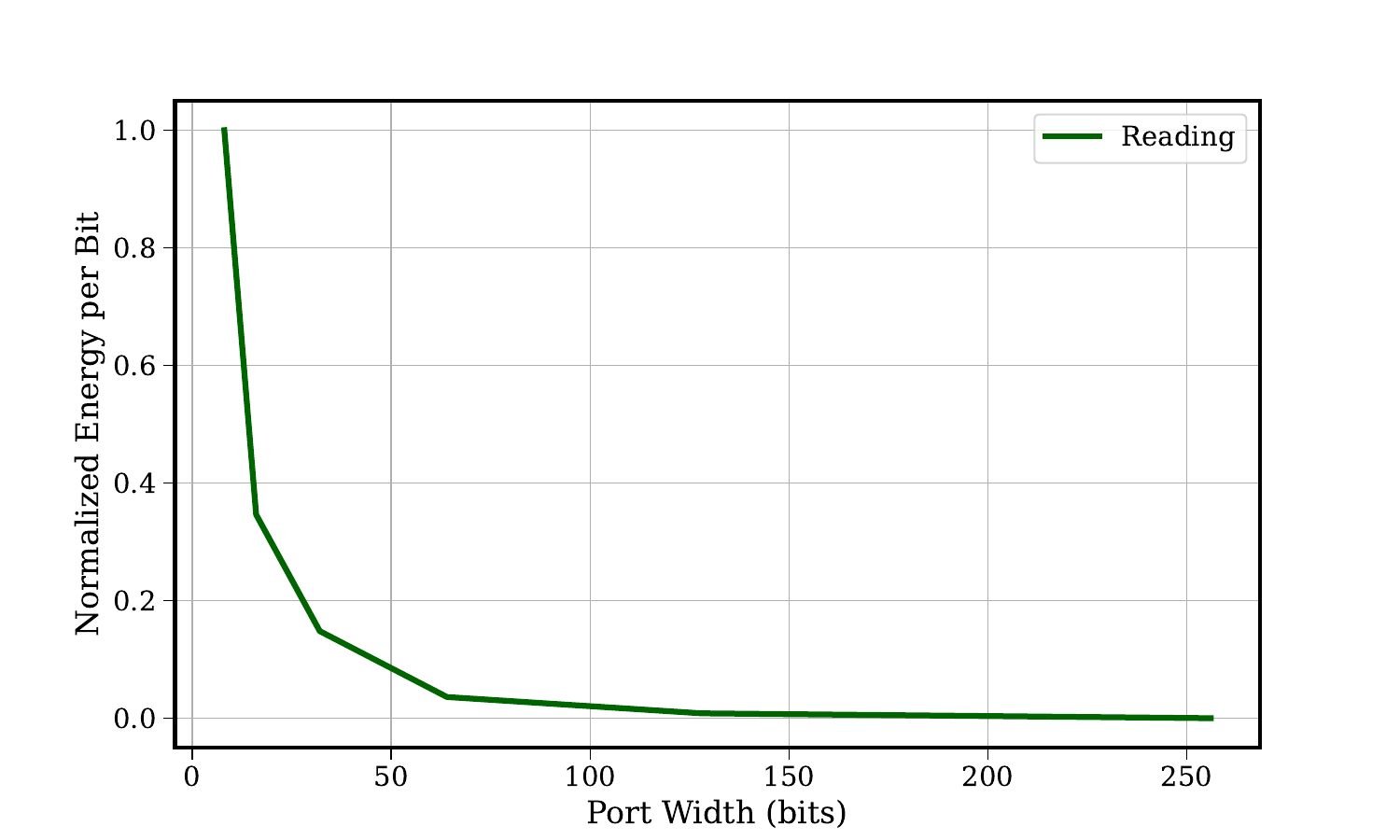}
    \caption{Normalized energy per bit loaded w.r.t. the port width}
    \label{fig:mem-portwidth}
\end{figure}

SparrowSNN hardware uses a 2-level memory design because memory operations dominate energy consumption; thus, SparrowSNN uses wide SRAM ports with local buffering to serve fine-grained consumers efficiently.  
Figure~\ref{fig:mem-portwidth} (normalized\footnote{For confidential reason, we are not able to provide the exact values}) shows energy/bit with respect to the port width: wider ports reduce energy/bit but require buffers to down-convert to the compute granularity. According to this energy analysis, we have designed both weight and activation SRAMs to use 128-bit ports. Weights are consumed as 8-bit values; activations/spike representations are consumed at 1\,bit (IF) or 2/4/8/16\,bits (SSF or ANN), configurable per layer. Local input/output buffers bridge the 128-bit SRAM bursts to these per-op widths, minimizing total memory energy while keeping utilization high.

To minimize redundant memory operations, Sparrow includes dedicated input and output spike buffers. The input spike buffer handles 128-bit chunks from activation memory or external inputs, supporting FIFO reads with variable bit-widths of 1, 2, 4, 8, or 16 bits. In SSF and ANN modes, bit-widths correspond to the ANN quantization levels or SSF time window size (e.g., $\log_2(T+1)$ bits for an SSF layer equivalent to a T-bit IF layer). Loaded input values are multiplied by the corresponding weights and accumulated in the Membrane Potential Buffer.

In IF mode, input spikes bypass direct addition when zero-valued, avoiding unnecessary computation but not skipping memory operations. To reduce costly redundant weight loading observed in typical timestep-first processing (illustrated in Figure~\ref{fig:redundant-weight-loading}), Sparrow buffers membrane potentials across multiple timesteps (up to 16) in a dedicated buffer, reducing memory accesses at the cost of additional buffering.

\begin{lstlisting}[float,caption=Inference Process Pseudo-code, label={lst:process},basicstyle=\footnotesize\ttfamily,keywordstyle=\bfseries, 
commentstyle=\itshape\color{gray}, 
morekeywords={for, in, as}]

for l in 1..|layers| :
    for n_out in  0..|layers[l]| :
        for n_in in 0..|layers[l-1]| :
            // Initialize mambrane potential
            V := 0;
            V_buf[0..T] := 0;
            // Input Accumulation
            if layers[l].type == SSF || ANN :
                V += W[n_in][n_out] * (ACT[n_in] as int8);
            else : // layers[l].type == IF
                for t in 0..T :
                    if ACT[n_in][t] == 1 :
                        V_buf[t] += W[n_in][n_out];
            // Bias
            if layer[l].type == IF:
                for t in 0..T :
                    V_buf[t] += B[l];
            else :// layer[l]. type == SSF || ANN
                V +=  B[l];
                        
            // Activation 
            if layers[l].type == IF
                for t i 0..T :
                    if V >= threshold[l] :
                        ACT[n_out][t] = 1; 
                        V -= threshold[l];
                    else :
                        V += V[t];
            else if layers[l].type == SSF:
                ACT[n_out] = V/threshold[l];
            else :// layers[l].type == ANN
                // ReLU
                V = MAX (0, V[0]);
                // Re-Quantization
                V = (V * Q) << Shift
                ACT[n_out] = V;
\end{lstlisting}

After processing all inputs for a neuron, the corresponding bias is loaded from weight memory, added to the membrane potential, and passed to the compare-and-spike unit. For SSF and IF layers, this unit compares the membrane potential against a threshold value retrieved from configuration registers and generates spikes.
For ANN layers, the unit implements a ReLU and quantization activation function. Generated spikes are temporarily stored in the output spike buffer until reaching 128 bits, at which point they are transferred to activation memory. Once all layers are processed, the last activation values are put into the classifier, which outputs the category of the classification based on the highest value of the last layer.

Listing~\ref{lst:process} illustrates the inference process as a pseudo-code. The FSM-based controller follows the processing order of this code to activate the components and to load weights/activation from memories by providing addresses and enable signals. Specifically, We apply two hardware optimizations aligned with SSF. First, instead of performing $T\!\cdot\!s$ per-synapse accumulator updates across timesteps, the controller aggregates per-input counts and issues a single MAC per synapse. This choice is compute-energy driven according to our discussion in around Table~\ref{tab:energy-comp}: an 8b-8b-16b MAC costs $\sim$2.6$\times$ an 8b-16b ACC. In our ECG/EEG setting ($T\!=\!31$, $s\!\ge\!0.2$ so $Ts\!\ge\!6$), one $\lceil\log_2(T{+}1)\rceil$-bit MAC per synapse is cheaper than $Ts$ accumulations; with SSF’s single weight read per window, total energy is lower. Second, for computing the neuron's output in SSF mode, after accumulating V, we directly compute the spike count as 
$c=\lfloor \frac{max(0,V)}{\theta} \rfloor$ (clipped to the window length), bypassing per-spike comparisons since only the total number of spikes is required.

\subsection{Handling the sparsity in SNN}
Sparsity in SNNs arises from activations that become zero at runtime. Leveraging this sparsity requires an additional step: dynamically detecting and skipping zero activations to avoid unnecessary computations.  To implement sparsity detection, the memory bus width must match the weight size (8-bit) to selectively read weights associated only with non-zero activations. Additionally, a zero-detection mechanism must be included in the computation unit (CU) to skip multiply-accumulate (MAC) operations for zero activations. Although this method reduces the theoretical computation count, it introduces two major overheads: 1) Zero-Detection Mechanism: An extra hardware component is required to detect zero activations, adding complexity and energy overhead. 2)Reduced Memory Bus Efficiency: Narrowing the memory bus to match the weight size significantly decreases bus efficiency, increasing energy consumption due to more frequent memory accesses. Our experimental results confirm that incorporating sparsity detection into our ultra-low-power design leads to an energy increase, primarily due to memory bus inefficiencies. Thus, we conclude that exploiting sparsity does not provide practical energy benefits and therefore do not include this mechanism in our final ultra-low-power implementation.

\section{Evaluations}
\label{sec:result}
\subsection{Experiment Setup}
In this study, we utilized CUDA-accelerated PyTorch version 1.12.1+cu116 for training with two NVIDIA A100 GPUs. We synthesized our ASIC design with a commercialized 22nm technology node and with memory compiler in low-power configurations. Due to the simplicity of our design, we developped a straightforward cycle-accurate simulation to validate our design and for performance evaluation. We derived the end-to-end energy consumption by combining the synthesis power reports (with explicit power analyses on the control, compute and memories components) and the active time of each component obtained from the cycle-accurate simulation.
The evaluated corner is at typical (TT) process, 0.85V, 25C temperature.

\subsubsection{Workflow}
We begin by applying NAS to optimize model architectures for edge signals. For example, NAS identifies a 5-layer architecture [32, 64, 32, 16, 64] for ECG classification on the MIT-BIH dataset, and a more compact 3-layer architecture [128, 32, 32] for EEG classification on the DEAP dataset. We then use these tailored architectures to construct a hybrid model, which is synthesized and deployed on our custom ASIC design. During the deployment phase, encoded signals can be directly processed by the hardware.

\subsubsection{Preprocessing of the datasets}
Aiming low-power scenarios, we target two common bio-medical applications: the electrocardiogram (ECG) classification task on the MIT-BIH dataset, and one electroencephalogram (EEG) classification task on the DEAP dataset. As recommended by AAMI~\cite{mar2011optimization}, due to significant imbalances in heart beat types within these records, we excluded four of the MIT-BIH dataset's total of 45 recods (numbers 120, 104, 107, and 217). Additionally, we removed the baseline and normalized the remaining records to the range \([0,1]\), following methods from prior research~\cite{mar2011optimization}. In this dataset, the R peaks are manually annotated; thus, we extracted a window of 180 data points centered around each R peak—90 points on each side—to segment the beats. To construct the training and inference dataset, we allocated 60\% of the heartbeats for training, 20\% for validation, and an additional 20\% for testing. Given the imbalance in the training dataset, we employed synthetic minority oversampling (SMOTE)~\cite{chawla2002smote} to achieve a balanced dataset. For EEG DEAP dataset processing, we split the 32-channel EEG signals into non-overlapping 2-second windows (256 samples per window) and filtered with a band-pass filter (4–45Hz) ~\cite{koelstra2011deap} to remove noise.
For each window, we extract several features: power spectral density (PSD) using Welch’s method ~\cite{welch1967use} over four standard frequency bands (theta: 4–8Hz, alpha: 8–14Hz, beta: 14–30Hz, gamma: 30–45Hz); differential entropy (DE) on the filtered signal~\cite{duan2013differential}; and Hjorth parameters to capture time-domain characteristics~\cite{oh2014novel}. The features are then normalized for each subject to reduce individual differences. We reduce the feature dimensions with principal component analysis (PCA)~\cite{marjit2021eeg} and add Gaussian noise for data augmentation.

\subsection{Inference on ECG signal classification}

\begin{figure}[h]
    \centering
    \includegraphics[width=1.0\linewidth]{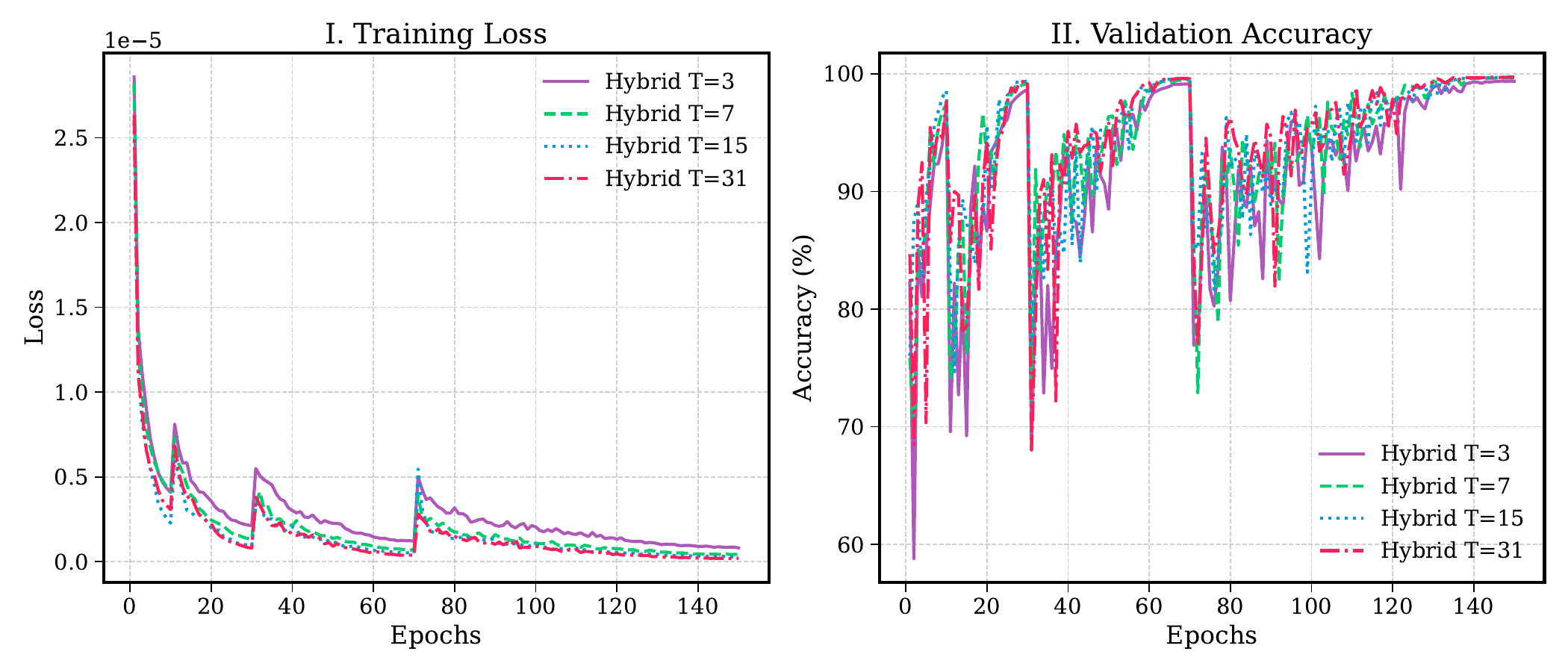}
    \caption{Training loss and validation accuracy for Hybrid model. }
    \label{fig: loss_ecg}
\end{figure}

We evaluated our IF, SSF and Hybrid ANN-SSF models on the MIT-BIH dataset for their training performance, accuracy, energy consumption, and compareed with existing state-of-the-art methods. Specifically, we trained each model for 150 epochs with an initial learning rate of 0.01, using an 60/20/20 training/validation/test split. The cosine annealing warm restarts scheduler (initial restart period \(T_0=10\), restart multiplier \(T_{mult}=2\), minimum learning rate \(\eta_{min}=1\times10^{-6}\)) is used to effectively mitigate overfitting and enhance convergence~\cite{nagubandi2022electric}. To verify the model is fully trained, We show the training loss and accuracy in Figure~\ref{fig: loss_ecg}. The decreasing loss and increasing validation accuracy shows the model is well-trained and not overfitting.

\begin{figure}[h]
    \centering
    \includegraphics[width=.8\linewidth]{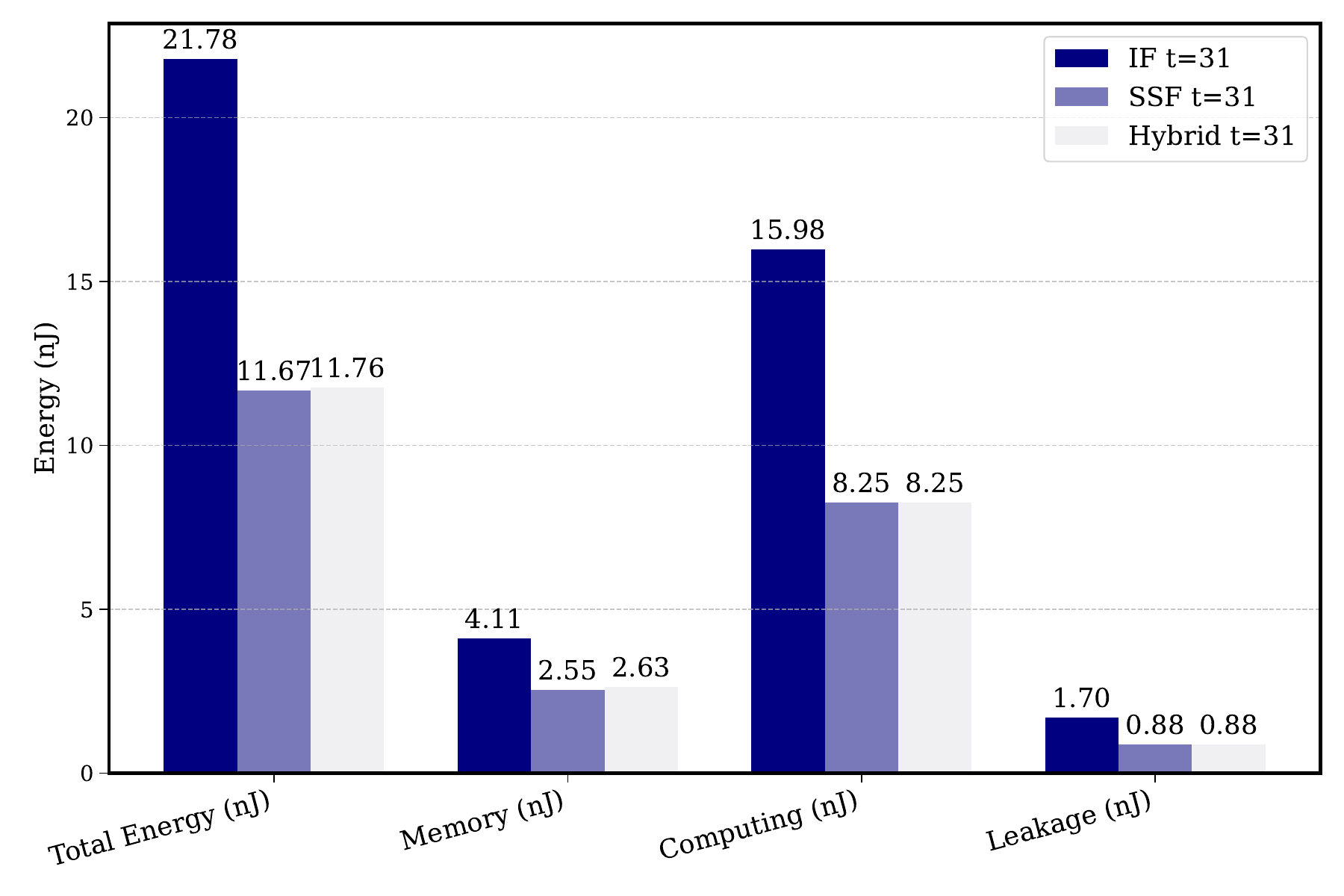}
    \caption{Energy consumption comparison for various model at T=31 for ECG dataset. }
    \label{fig:ecg_result_t31}
\end{figure}

\begin{table}[h]
\centering
\caption{Comparison of various models on MIT-BIH ECG classification dataset.}
\label{tab:snnalays}
\resizebox{\linewidth}{!}{
\begin{tabular}{c|ccc|ccc|ccc}
\hline
     & \multicolumn{3}{c|}{SNN with IF} & \multicolumn{3}{c|}{SNN with SSF} & \multicolumn{3}{c}{Hybrid Model} \\ \cline{2-10} 
     & \begin{tabular}[c]{@{}c@{}}Acc\\ (\%)\end{tabular}
     & \begin{tabular}[c]{@{}c@{}}Energy\\ (nJ)\end{tabular}
     & \begin{tabular}[c]{@{}c@{}}\rev{Latency}\\ \rev{(ms)}\end{tabular}
     & \begin{tabular}[c]{@{}c@{}}Acc\\ (\%)\end{tabular}
     & \begin{tabular}[c]{@{}c@{}}Energy\\ (nJ)\end{tabular}
     & \begin{tabular}[c]{@{}c@{}}\rev{Latency}\\ \rev{(ms)}\end{tabular}
     & \begin{tabular}[c]{@{}c@{}}Acc\\ (\%)\end{tabular}
     & \begin{tabular}[c]{@{}c@{}}Energy\\ (nJ)\end{tabular}
     & \begin{tabular}[c]{@{}c@{}}\rev{Latency}\\ \rev{(ms)}\end{tabular} \\ \hline
T=3  & 73.35 & 12.12 & \rev{0.132} & 87.42 & 11.49 & \rev{0.124} & 97.97 & 11.66 & \rev{0.124} \\
T=7  & 86.8  & 13.50 & \rev{0.148} & 92.48 & 11.55 & \rev{0.124} & 97.89 & 11.69 & \rev{0.124} \\
T=15 & 89.66 & 16.26 & \rev{0.179} & 97.03 & 11.61 & \rev{0.124} & 98.56 & 11.72 & \rev{0.124} \\
T=31 & 84.15 & 21.78 & \rev{0.241} & 97.53 & 11.67 & \rev{0.124} & 98.61 & 11.76 & \rev{0.124} \\ \hline
\end{tabular}}
\end{table}
As shown in Table~\ref{tab:snnalays} and Figure~\ref{fig:ecg_result_t31}, the Hybrid ANN-SSF model achieved higher validation accuracy (98.61\% at $T=31$ compared to the SSF model alone (97.53\%), suggesting the inclusion of ANN layers enhances feature extraction and classification accuracy. Further, from an energy perspective, at $T=31$, the Hybrid model consumes only 11.76 nJ per inference, slightly higher than SSF (11.67 nJ) but significantly lower than the IF model (21.78 nJ).

\begin{table*}[hbt]
\centering
\caption{Comparison with SOTA related works on MIT-BIH ECG classification dataset.}
\label{tab:sota-comp-ecg}
\resizebox{\linewidth}{!}{
\begin{tabular}{c|ccccccc}
\hline
                 & OJCAS21~\cite{chuang2021arbitrarily} & AS23~\cite{zhang2023low} & TCAS-II21~\cite{cai20211} & TBioCAS22~\cite{mao2022ultra} & ISSCC22~\cite{liu202282nw}  & TBioCAS22~\cite{chu2022neuromorphic} & This works \\ \hline
Process (nm)     & 40                 & 55         & 180    & 28         & 180          & 40                & 22      \\
Voltage (V)            & 0.6                   & 1.2          & 1.8      & 0.57         & 1.2          & 1.1                 & 0.8        \\
Frequency (Hz)       & 70M                   & 100K         & 5M       & 250k-5M      & Clock-Free & 65k-100M            & 10-100M         \\
Areas (mm2)      & 0.2123                & 0.33         & 0.75     & 0.54         & 10           & 0.3246              & 0.114     \\
Energy/inference & 0.212$\mu$J                 & 0.234$\mu$J      & 1.8$\mu$J    & 0.3$\mu$J        & N/A          & 0.75$\mu$J              & 0.01176$\mu$J    \\
Methods          & ELM                   & ANN          & ANN      & SNN          & SNN          & SNN                 & SNN        \\
Accuracy (\%)    & 92                    & 96.69        & 98       & 98.6         & 90.5         & 98.22               & 98.61     \\ \hline
\end{tabular}}
\end{table*}

\begin{figure}[h]
    \centering
\includegraphics[width=\linewidth]{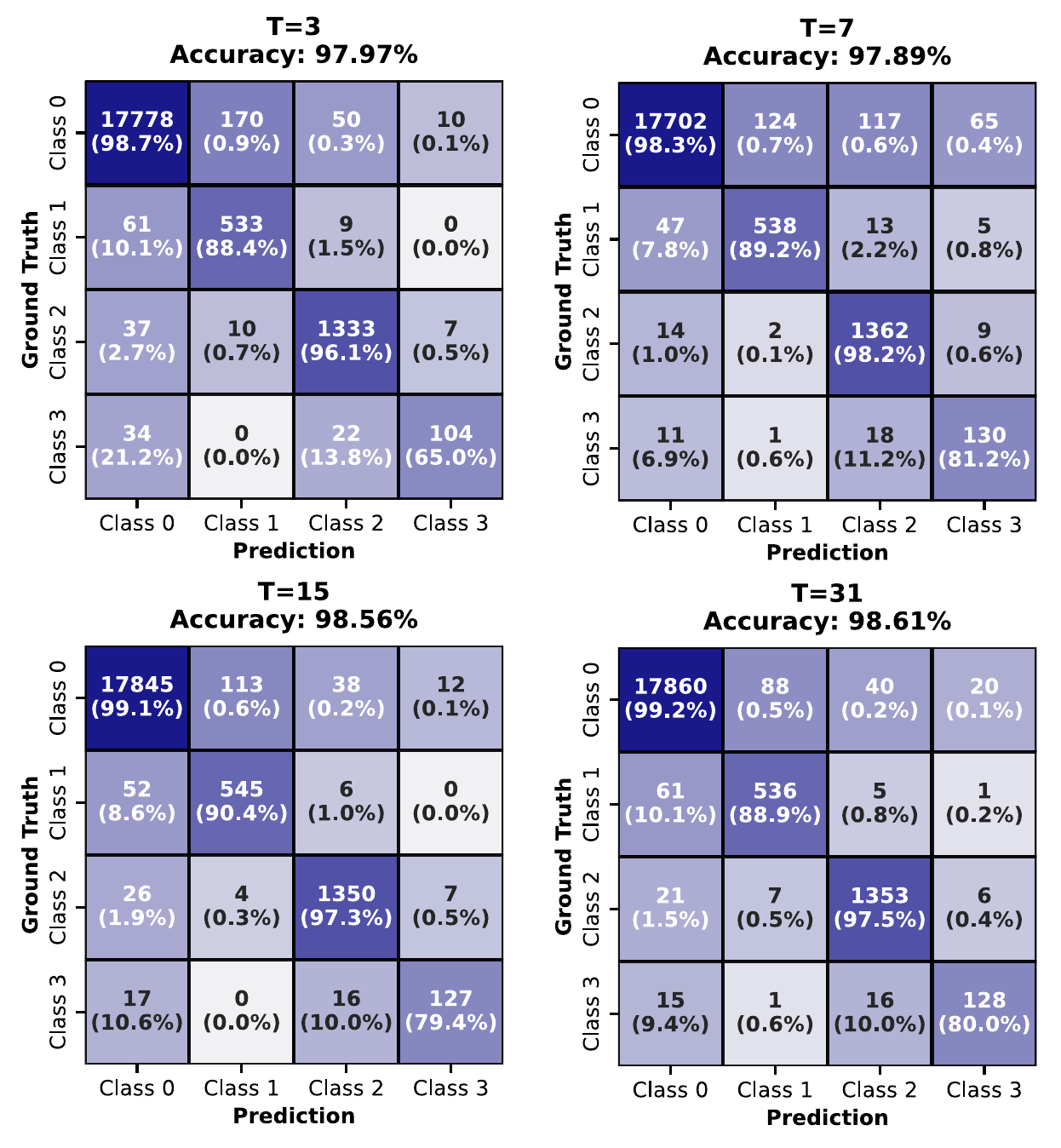}
    \caption{Confusion matrix for Hybrid model. }
    \label{fig:cm_ecg}
\end{figure}

We further show the confusion matrix for the Hybrid model with different timesteps to demonstrate the classification accuracy on the unbalanced testing dataset, where most of the heartbeats are normal (class 0). As we can see from Figure~\ref{fig:cm_ecg}, each label from 0-3 is well classified with T=31, achieving 99.2\% accuracy for class 0 and 96.5\% for class 2.

%\textit{Training loss and validation accuracy} of the SSF and Hybrid models. We utilize the cosine annealing warm restarts scheduler (initial restart period \(T_0=10\), restart multiplier \(T_{mult}=2\), minimum learning rate \(\eta_{min}=1\times10^{-6}\)) to effectively mitigate overfitting and enhance convergence.  \textbf{c}\textit{ Energy consumption breakdown.} (Memory, Computation, and Leakage energy) comparing IF, SSF, and Hybrid models at an identical spike window size (\(T=31\)) for fair evaluation.   \textbf{d/e)}\textit{ Confusion matric} for SSF-based SNN and Hybrid models across different spike time windows, demonstrating model robustness and accuracy improvements. \textbf{f)}\textit{Comparison of proposed models} against state-of-the-art ECG classification methods, highlighting superior accuracy and significantly lower energy consumption.

In this work, we benchmark our design against recently published state-of-the-art (SOTA) ECG classification accelerators. Chu {\em et al.}~\cite{chu2022neuromorphic} introduced a spike-driven SNN processor for always-on ECG classification that leverages level-crossing (LC) sampling for efficient temporal coding and event-driven processing. Combined with hardware-aware STBP training, their design achieves 98.22\% accuracy with 0.75$\mu$J energy per inference, which is 7.72\% higher accuracy than reported by Liu {\em et al.}~\cite{liu202282nw}. Our ASIC achieves substantially lower energy (11.76\,nJ per inference) while maintaining high detection accuracy (98.61\%). We also compare against leading ANN-based designs; for example, Zhang {\em et al.}~\cite{zhang2023low} reports 96.69\% accuracy with 234.4\,nJ per inference. We emphasize energy-per-inference rather than power because operating frequencies and measurement conditions differ across platforms, making power comparisons less directly comparable.
\rev{Further, process-node labels do not translate cleanly into predictable performance/energy scaling: realized PPA gains vary across foundries and are strongly affected by DTCO and other design choices, which complicates attributing cross-platform differences to technology node alone. measured cross-generation studies show that microarchitectural and system-level choices can dominate power/performance outcomes~\cite{itrs,esmaeilzadeh2011looking,esmaeilzadeh2011dark}. Accordingly, while technology-node differences may contribute, the large gains observed for sparrowSNN are unlikely to be explained by node scaling alone.}

\begin{table*}[h]
\centering
\caption{Comparison with SOTA related works on DEAP EEG classification dataset.}
\label{tab:snnalays_eeg}
\resizebox{\linewidth}{!}{
\begin{tabular}{c|cccccc}
\hline
                              & ISCAS19~\cite{8702738}    & JEmSelTopC19~\cite{fang2019development} & TBioCAS 21 ~\cite{aslam202110}      & TBioCAS 20~\cite{aslam2020chip}  & JSEN24~\cite{martis2024low}  & This work \\ \hline
Process (nm)                  & 65         & 28           & 16               & 180         & 22      & 22                \\
Voltage (V)                   & /          & 0.9          & 1                & 1           & /       & 0.8               \\
Frequency (Hz)                & 1K         & 70M          & 200M             & 1K          & 10K     & 100M MAX           \\
Areas (mm2) & /          & 3.35         & 16               & 6           & 16      & 0.114             \\
Energy/Inference              & 10uJ       & 0.78uJ       & 10.13uJ          & 16uJ        & 0.417uJ & 0.01787uJ          \\
Methods                       & SVM        & CNN          & MLP              & SVM         & SNN     & SNN               \\
Accuracy (Valence/Arousal\%)    & 63/60 & 83.36/76.67 & 85.4 (4 classes) & 72.96/73.14 & /       & 85.04/85.31       \\ \hline
\end{tabular}}
\end{table*}

% \begin{figure*}[t]
%     \centering
%     \includegraphics[width=1.0\linewidth]{figures/test_ecg.pdf}
%     \caption{ECG classification based on MIT-BIH dataset. }
%     \label{fig:ecg_result}
% \end{figure*}

\subsection{Inference on EEG signal classification}

\begin{figure}[h]
\begin{subfigure}{\linewidth}
     \centering
\includegraphics[width=1.0\linewidth]{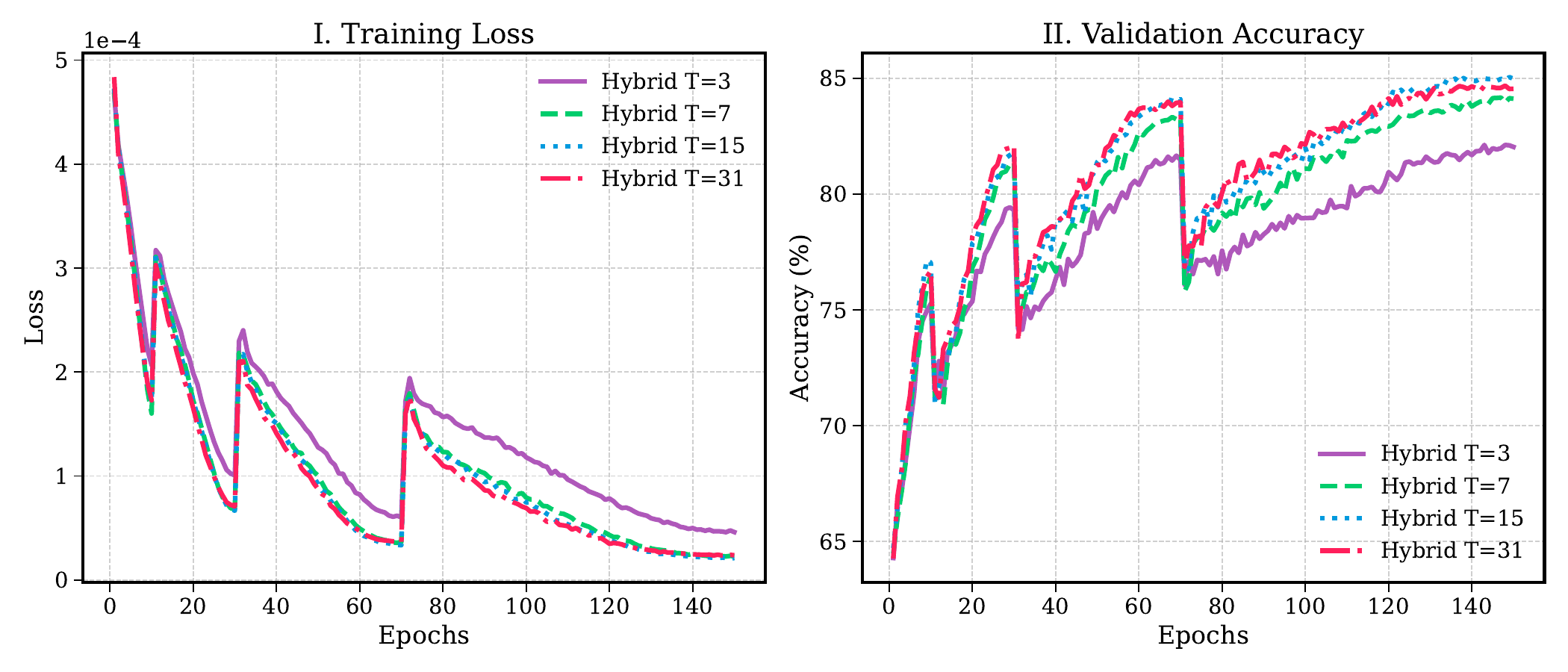}
    \caption{Training loss and validation accuracy for Hybrid model (valence dimension). }
    \label{fig:eeg_va_tr}
\end{subfigure}

\begin{subfigure}{\linewidth}
    \centering
\includegraphics[width=1.0\linewidth]{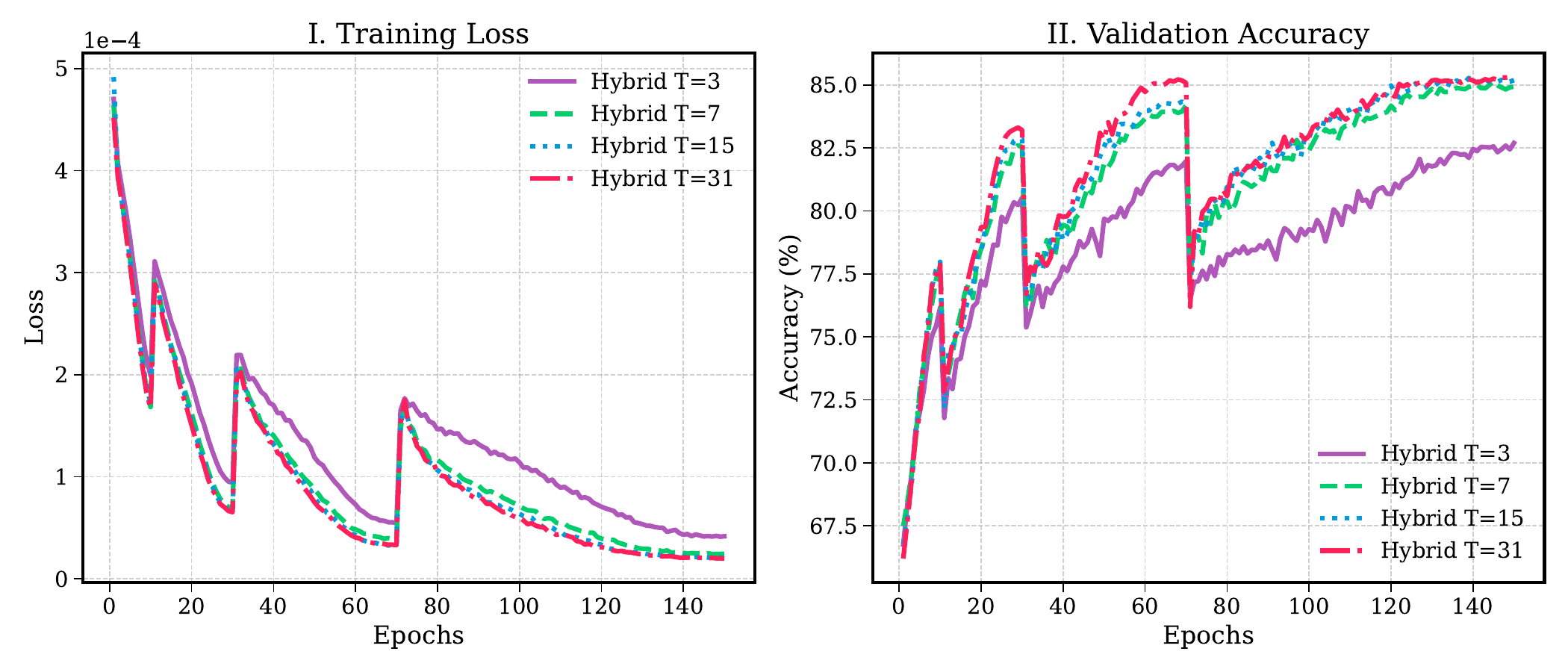}
    \caption{Training loss and validation accuracy for Hybrid model (arousal dimension). }
        \label{fig:eeg_ar_tr}
\end{subfigure}
\end{figure}

We conducted extensive experiments on the DEAP dataset to assess the effectiveness of our proposed SSF-based SNN and Hybrid ANN-SNN architectures for EEG-based valence and arousal classification. Our evaluation emphasizes training stability, accuracy performance, energy efficiency, and a detailed comparison against previously reported state-of-the-art EEG classification architectures. We first show the training loss and accuracy for both valence dimension and arousal dimension respectively in Figure~\ref{fig:eeg_va_tr} and Figure~\ref{fig:eeg_ar_tr}.

\begin{figure}[h]
    \centering
    \includegraphics[width=.8\linewidth]{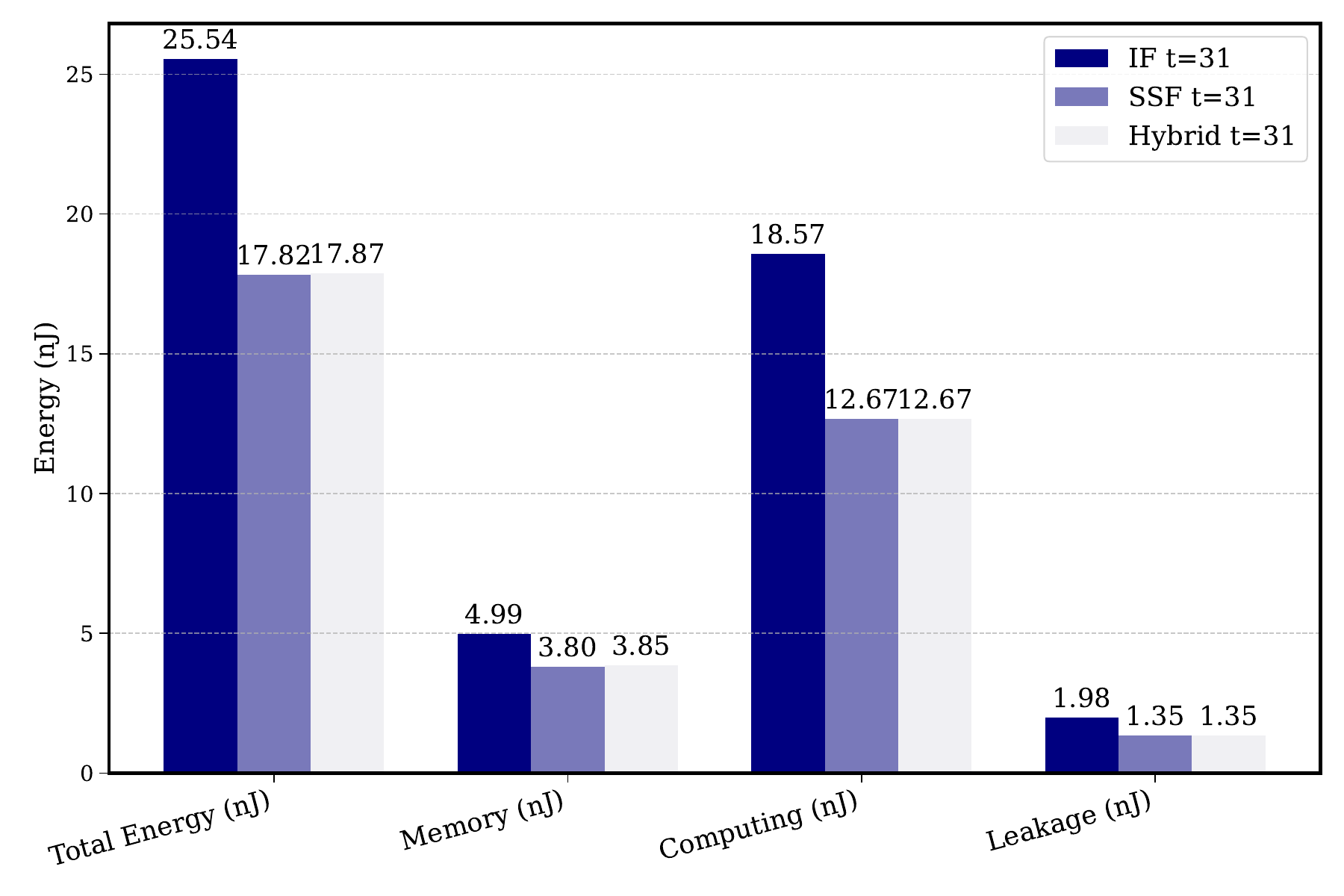}
    \caption{Energy consumption comparison for various model at T=31 for EEG dataset. }
    \label{fig:eeg_result_t31}
\end{figure}
\begin{figure}[h]
    \centering
    \begin{minipage}[t]{\linewidth}
        \centering
        \includegraphics[width=.85\linewidth]{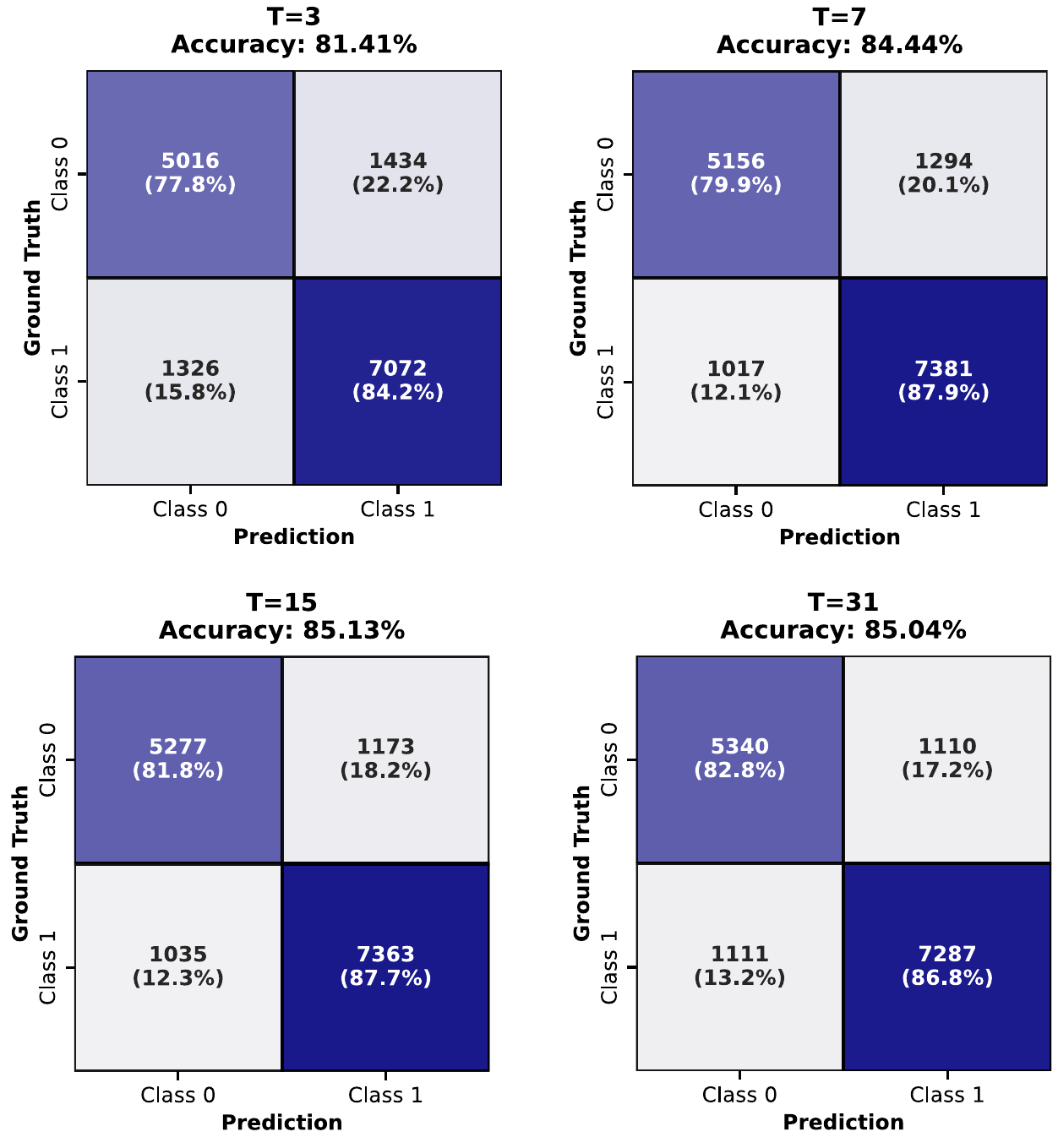}
        \caption{Hybrid model confusion matrix (valence dimension).}
        \label{fig:valence_confusion}
    \end{minipage}
    \hfill
    \begin{minipage}[t]{\linewidth}
        \centering
        \includegraphics[width=.85\linewidth]{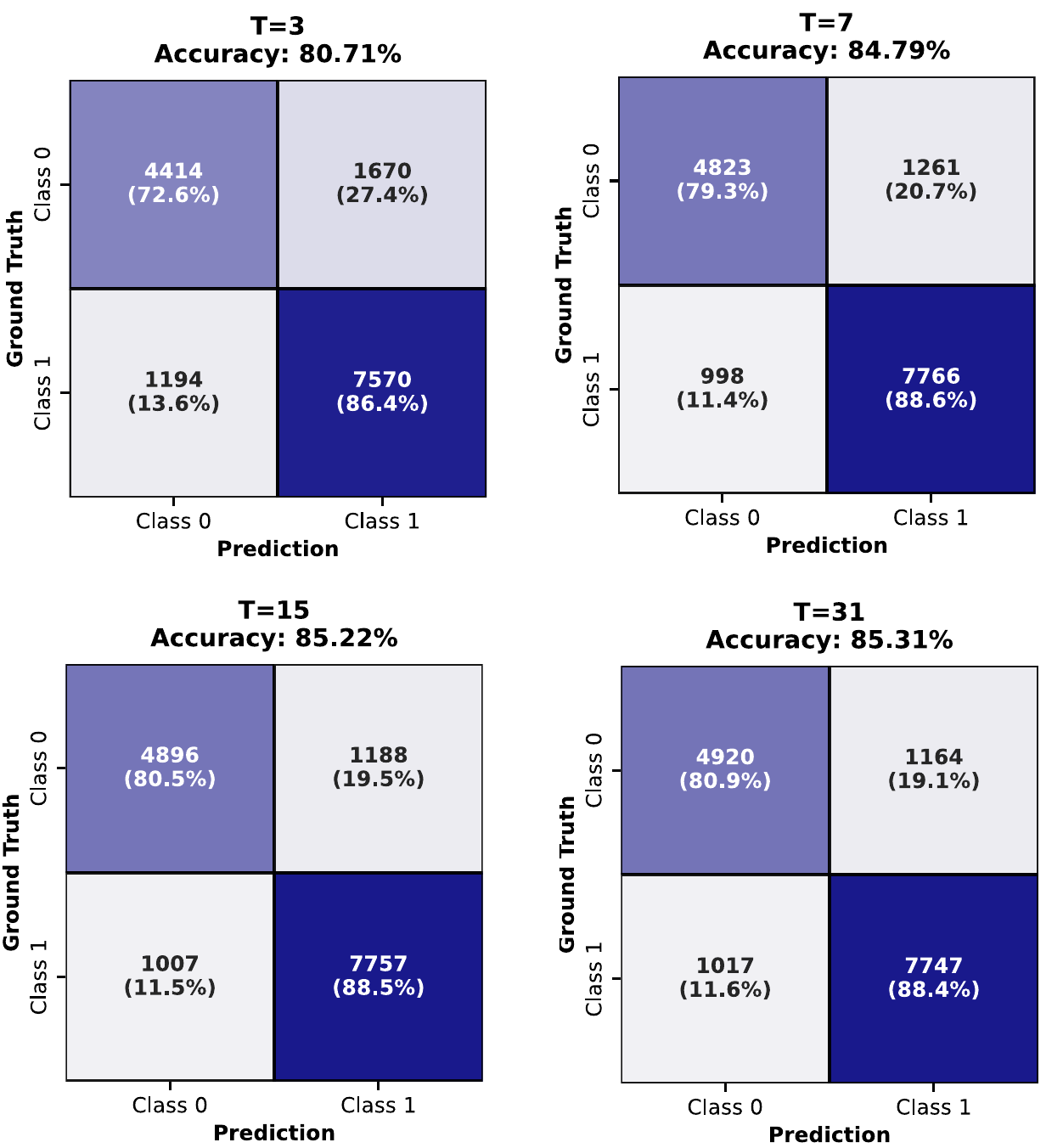}
        \caption{Hybrid model confusion matrix (arousal dimension).}
        \label{fig:arousal_confusion}
    \end{minipage}
\end{figure}

As shown in Table~\ref{tab:eeg_snnalays} and Figure~\ref{fig:eeg_result_t31}, our energy evaluation demonstrates clear advantages for our proposed neuron designs. Specifically, at a representative spike window size of T=31, the Hybrid model achieves high accuracy (85.04\% valence, 85.31\% arousal) with a modest energy consumption of 17.87 nJ per inference. This is only slightly higher than the SSF neuron alone (17.82 nJ per inference), yet substantially lower than the traditional IF model (25.54 nJ). The minimal additional energy overhead of the Hybrid model relative to SSF alone is well justified by its accuracy improvement. The confusion matrix for arousal and valance dimension is shown in Figure ~\ref{fig:valence_confusion} and ~\ref{fig:arousal_confusion}.

\begin{table}[t]
\centering
\caption{Comparison of various models on DEAP EEG classification dataset (V indicates valence and A indicates arousal).}
\label{tab:eeg_snnalays}
\scalebox{0.67}{\begin{tabular}{c|ccc|ccc|ccc}
\hline
     & \multicolumn{3}{c|}{SNN with IF} & \multicolumn{3}{c|}{SNN with SSF} & \multicolumn{3}{c}{Hybrid Model} \\ \cline{2-10} 
     & \begin{tabular}[c]{@{}c@{}}Accuracy\\ (V/A\%)\end{tabular}
     & \begin{tabular}[c]{@{}c@{}}Energy\\ (nJ)\end{tabular}
     & \begin{tabular}[c]{@{}c@{}}\rev{Latency}\\ \rev{(ms)}\end{tabular}
     & \begin{tabular}[c]{@{}c@{}}Accuracy\\ (V/A\%)\end{tabular}
     & \begin{tabular}[c]{@{}c@{}}Energy\\ (nJ)\end{tabular}
     & \begin{tabular}[c]{@{}c@{}}\rev{Latency}\\ \rev{(ms)}\end{tabular}
     & \begin{tabular}[c]{@{}c@{}}Accuracy\\ (V/A\%)\end{tabular}
     & \begin{tabular}[c]{@{}c@{}}Energy\\ (nJ)\end{tabular}
     & \begin{tabular}[c]{@{}c@{}}\rev{Latency}\\ \rev{(ms)} \end{tabular} \\ \hline
T=3  & 55.54/58.24 & 18.17 & \rev{0.197} & 70.90/70.95 & 17.69 & \rev{0.191} & 81.41/80.71 & 17.78 & \rev{0.191} \\
T=7  & 55.21/56.85 & 19.22 & \rev{0.209} & 75.66/76.27 & 17.73 & \rev{0.191} & 84.44/84.79 & 17.81 & \rev{0.191} \\
T=15 & 63.95/65.83 & 21.33 & \rev{0.232} & 79.14/79.74 & 17.78 & \rev{0.191} & 85.13/85.22 & 17.84 & \rev{0.191} \\
T=31 & 71.79/73.43 & 25.54 & \rev{0.280} & 82.64/83.02 & 17.82 & \rev{0.191} & 85.04/85.31 & 17.87 & \rev{0.191} \\ \hline
\end{tabular}}
\end{table}

We further compared our proposed designs against recent state-of-the-art EEG classification works. Abdul Rehman Aslam et al. ~\cite{aslam202110} introduced an SVM-based EEG classification processor with hardware-efficient features, achieving accuracies of 63\% (valence) and 60\% (arousal) at an energy cost of 10 $\mu$J per inference using 65 nm CMOS. Another SVM implementation by Aslam ~\cite{aslam2020chip}improved performance (72.96\% valence, 73.14\% arousal) but required higher energy (16 $\mu$J per inference). CNN-based designs such as Fang et al.~\cite{fang2019development} attained higher accuracies (76.67\% valence, 83.36\% arousal) with 0.78 $\mu$J per inference at 28 nm, while Gonzalez’s BioCNN ~\cite{gonzalez2020inference} reached accuracies of 83.12\% (valence) and 76.78\% (arousal), but consumed significantly more energy (139.5 $\mu$J). Lastly, Luca Martis's FPGA-based SNN model~\cite{martis2024low} reported a low energy of 0.417 $\mu$J per inference but did not specify accuracy. Compared to these methods, our SSF-based SNN achieves higher accuracy (82.64\% valence, 83.02\% arousal) with substantially lower energy (0.01782 $\mu$J per inference). Our Hybrid ANN-SSF model further improves accuracy to 85.04\% (valence) and 85.31\% (arousal) at a comparable minimal energy cost (0.01787 $\mu$J per inference). This analysis confirms that our Hybrid model offers a better balance of accuracy and energy efficiency compared to existing state-of-the-art EEG classification architectures.

\section{Conclusion}

In this paper, we explored the accuracy and energy efficiency of SNNs from a hardware perspective and introduced SparrowSNN, a fully co-designed software-hardware ASIC optimized for ultra-low-power edge applications. We proposed the sum-spikes-and-fire neuron, which outperforms conventional IF neurons in both accuracy and energy efficiency. Additionally, we developed a Hybrid ANN-SSF model, integrating ANN layers to enhance feature extraction while maintaining minimal energy overhead. Using Neural Architecture Search, we optimized network structures for ECG and EEG classification tasks, achieving state-of-the-art results. Our Hybrid model reached 98.61\% accuracy with 11.76 nJ per inference on the MIT-BIH ECG dataset and 85.04\%/85.31\% accuracy with 17.87 nJ per inference for EEG valence/arousal classification on DEAP. Extensive benchmarking against relate works confirms that SparrowSNN offers the best trade-off between accuracy and energy efficiency. By integrating algorithmic and hardware optimization, our work provides deeper insights into the energy efficiency and operational workflows of SNNs on hardware platforms, paving the way for advancements in ultra-low-power wearable AI applications.

\bibliography{sample-base}
\bibliographystyle{IEEEtran}

\begin{IEEEbiography}[{\includegraphics[width=1in,height=1.25in,clip,keepaspectratio]{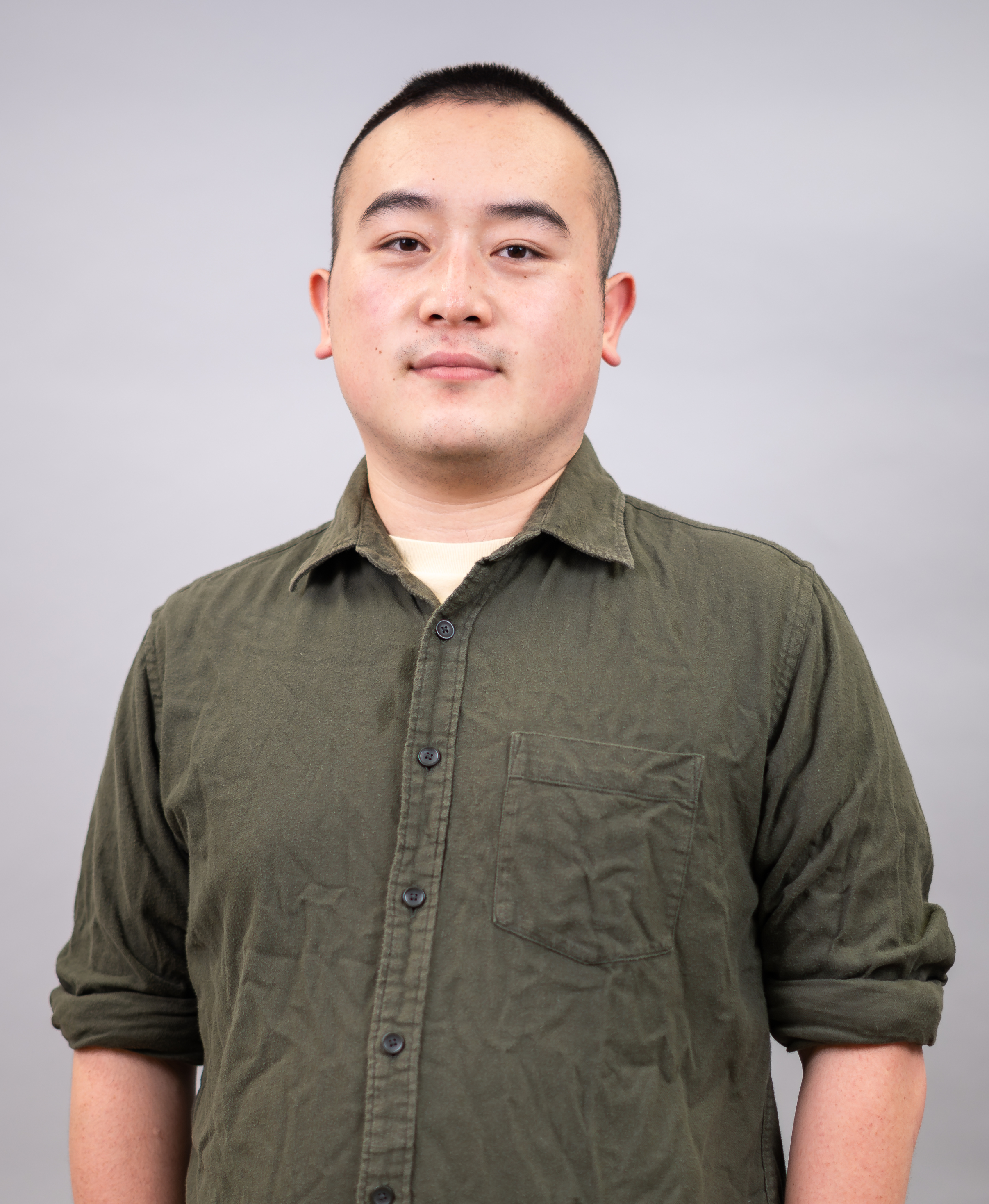}}]{Zhenyu Bai} is an ARTIC Fellow in the Department of Computer Science, National University of Singapore. His research interests include spatial dataflow architectures design and compilation techniques.
\end{IEEEbiography}

\begin{IEEEbiography}[{\includegraphics[width=1in,height=1.25in,clip,keepaspectratio]{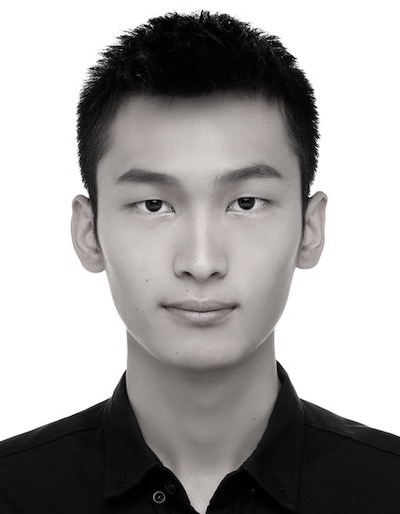}}]{Zhanglu Yan} received the BSc degree in Computer Science from Xi'an Jiaotong University in 2019, the MSc degree in Artificial Intelligence from the National University of Singapore (NUS) in 2020, and the PhD degree in Computer Science from NUS in 2024. He is currently a research fellow at NUS. His research focuses on neuromorphic computing and spiking neural networks.
\end{IEEEbiography}

\begin{IEEEbiography}[{\includegraphics[width=1in,keepaspectratio]{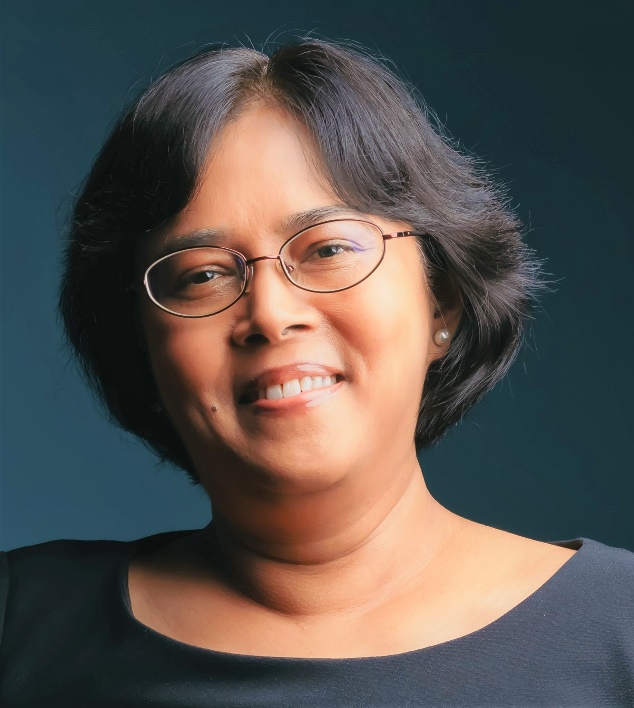}}]
{Tulika Mitra} is Dean, School of Computing and Provost’s Chair Professor of Computer Science at the National University of Singapore. Her research focuses on the hardware-software co-design of smart, energy-efficient, safety-critical embedded computing systems. 
\end{IEEEbiography}

\begin{IEEEbiography}[{\includegraphics[width=1in,height=1.25in,clip,keepaspectratio]{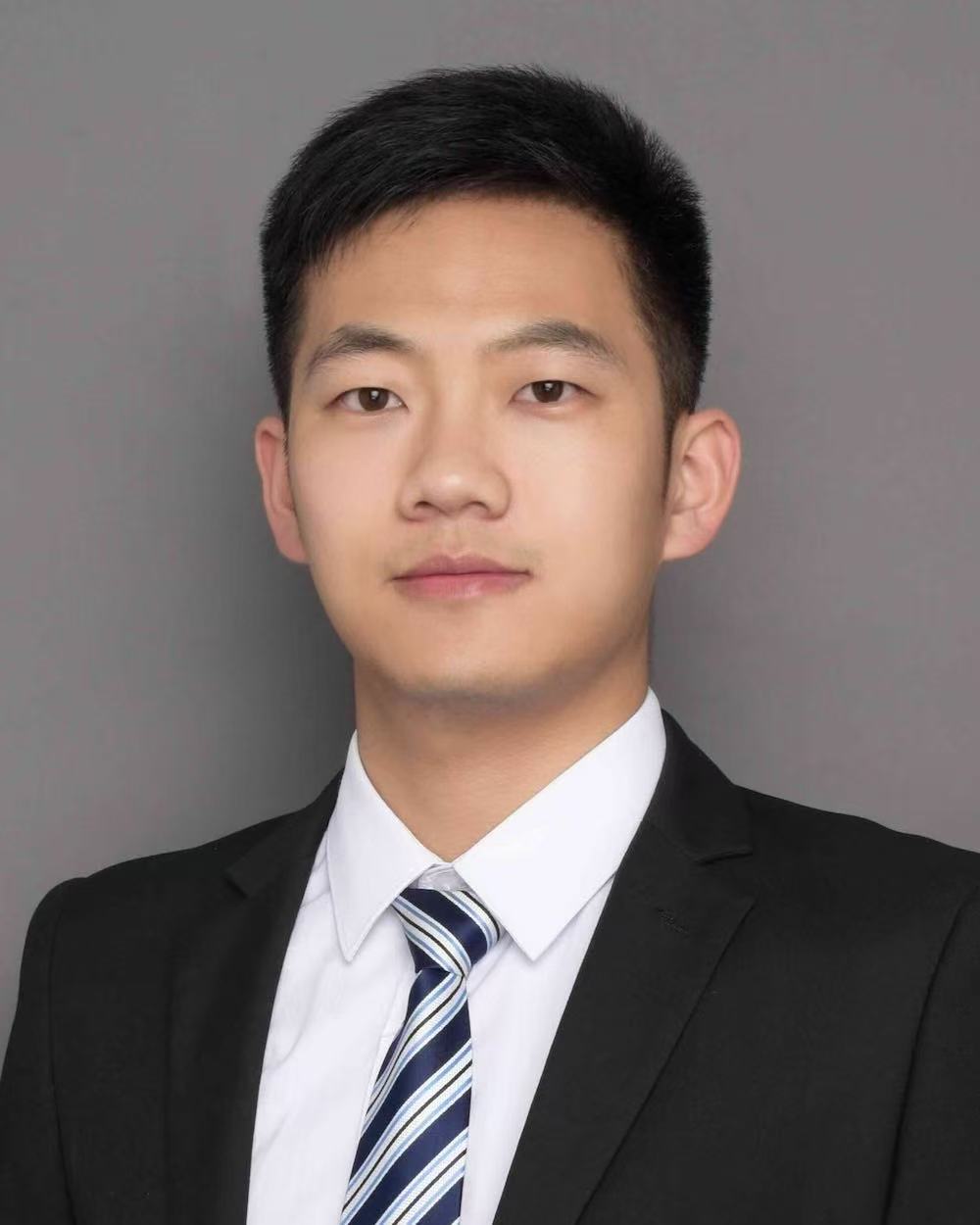}}]
{Bin Gao} received his bachelor's and master's degrees from Huazhong University of Science and Technology, China, in 2017 and 2020, respectively. He received his PhD from the National University of Singapore in 2025. His research interests include machine learning systems.
\end{IEEEbiography}

\begin{IEEEbiography}[{\includegraphics[width=1in,height=1.25in,clip,keepaspectratio]{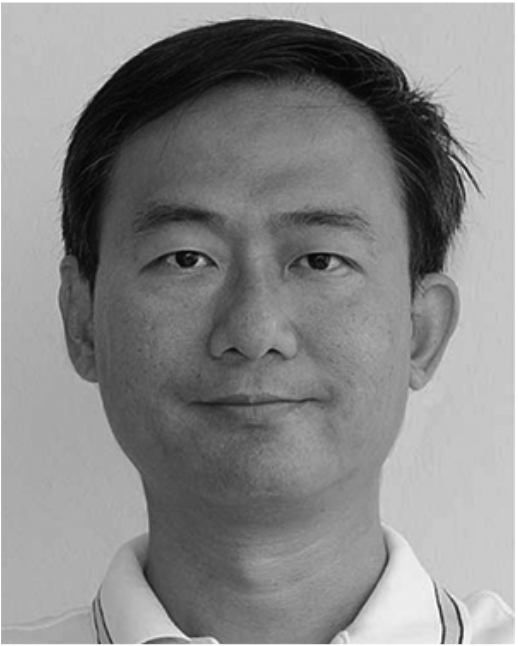}}]{Wong Weng-Fai} received the BSc degree from the National University of Singapore, in 1988, and the DrEngSc degree from the University of Tsukuba, Japan, in 1993. He is currently an associate professor with the Department of Computer Science, National University of Singapore. His research interests include computer architecture, compilers, and high-performance computing. He is a senior member of the IEEE.
\end{IEEEbiography}

\end{document}